\documentclass[%
 reprint, 
superscriptaddress,
 amsmath,amssymb,
 aps, physrev,
]{revtex4-2}
\usepackage{aas_macros}
\usepackage{graphicx}
\usepackage{dcolumn}
\usepackage{bm}
\usepackage{hyperref}
\usepackage[mathlines]{lineno}

\usepackage{color}
\definecolor{mypurple}{rgb}{0.5, 0, 0.85}
\definecolor{Hbetta}{rgb}{0,0.92,1}
\definecolor{myblue}{rgb}{0, 0.2, 0.85}
\definecolor{cadmiumgreen}{rgb}{0.0, 0.42, 0.24}
\definecolor{gold}{rgb}{0.7176, 0.5843, 0.0431}
\definecolor{org}{RGB}{33, 47, 61}

\begin{document}

\preprint{APS/123-QED}

\title{Low-Frequency Gravitational Waves in \\ Three-Dimensional Core-Collapse Supernova Models}

\author{Colter J. Richardson}
\email{cricha80@vols.utk.edu}
\affiliation{Department of Physics and Astronomy, University of Tennessee, Knoxville, TN 37996, USA}
\author{Anthony Mezzacappa}
\affiliation{Department of Physics and Astronomy, University of Tennessee, Knoxville, TN 37996, USA}
\author{Kya Schluterman}
\affiliation{Embry-Riddle Aeronautical University, 3700 Willow Creek Road, Prescott, Arizona 86301, USA}
\affiliation{Department of Physics and Astronomy, University of Tennessee, Knoxville, TN 37996, USA}
\author{Haakon Andresen} 
\affiliation{ The Oskar Klein Centre, Department of Astronomy, AlbaNova, SE-106 91 Stockholm, Sweden}
\author{Eric J. Lentz}
\affiliation{Department of Physics and Astronomy, University of Tennessee, Knoxville, TN 37996, USA}
\affiliation{Physics Division, Oak Ridge National Laboratory, P.O. Box 2008, Oak Ridge, Tennessee 37831-6354, USA}
\author{Pedro Marronetti}
\affiliation{Physics Division, National Science Foundation, Alexandria, Virginia 22314, USA}
\author{R. Daniel Murphy}
\affiliation{Department of Physics and Astronomy, University of Tennessee, Knoxville, TN 37996, USA}
\author{Michele Zanolin}
\affiliation{Embry-Riddle Aeronautical University, 3700 Willow Creek Road, Prescott, Arizona 86301, USA}

\date{\today}

\begin{abstract}    
    We discuss the low-frequency gravitational wave signals from three state-of-the-art three-dimensional core-collapse supernova models produced with the \textsc{Chimera} supernova code. We provide a detailed derivation of the gravitational wave signal sourced from the anisotropic emission of neutrinos and provide the total (fluid sourced and neutrino sourced) gravitational waves signal generated in our models\footnote{The data presented here can be found  at \href{https://doi.ccs.ornl.gov/dataset/847fc720-6ff7-50eb-a747-12fbb23038db}{Constellation: Chimera D-Series Gravitational Wave Emission Sourced from Neutrino Anisotropy}}. We discuss the templatablity of this low-frequency signal, which is useful for future work involving matched filtering for signal detection and parameter estimation. 
\end{abstract}

\maketitle
\section{Introduction}
\label{sec:Introduction}

    The detection of gravitational waves from a core-collapse supernova is an eventuality. In recent years, the advances in models of gravitational waves from core-collapse supernovae, the upgrades to and proposals for next-generation gravitational wave detectors, and the advancement of detection algorithms have propelled core-collapse supernova gravitational wave astronomy forward. Each of the so-called ``messengers'' emitted by a core-collapse supernova---gravitational waves, neutrinos, and photons---details different dynamics from different regions in the collapsing and exploding massive star. 
    Photons bring information from throughout the turbulent ejecta, neutrinos bring information from the neutrinospheres deep within the explosion, around the surface of the newly-formed proto- neutron star (PNS), and gravitational waves bring information from the very center of the explosion, outward. 
    
    Investigations of gravitational wave emission from core-collapse supernovae are often focused on high-frequency emission $\gtrapprox$ 500 Hertz, associated with oscillations within the PNS, excited by convection within it and accretion onto it
    \cite{1982A&A...114...53M, 1991A&A...246..417M, 1997A&A...317..140M, 2001ApJ...560L.163D, 2003PhRvD..68d4023K, 2004PhRvD..69l4004K, 2004ApJ...603..221M, 2009ApJ...697L.133K, 2009A&A...496..475M, 2009ApJ...707.1173M, 2010A&A...514A..51S, 2011ApJ...736..124K, 2012A&A...537A..63M, 2013ApJ...779L..18C, 2013ApJ...766...43M, 2013ApJ...768..115O, 2014PhRvD..89d4011K, 2015PhRvD..92h4040Y, 2015PhRvD..92l2001H, 2016PhRvL.116o1102H, 2016ApJ...829L..14K, 2017ApJ...851...62K, 2017PhRvD..95f3019R, 2017MNRAS.468.2032A, 2018ApJ...861...10M, 2018ApJ...865...81O, 2018MNRAS.475L..91T, 2018MNRAS.477L..96H, 2018ApJ...867..126K, 2018ApJ...857...13P, 2019MNRAS.486.2238A, 2019ApJ...876L...9R, 2019MNRAS.482..351V, 2019PhRvD.100d3026S, 2019MNRAS.487.1178P, 2020PhRvD.102b3027M, 2020MNRAS.494.4665P, 2020MNRAS.493L.138S, 2020ApJ...898..139W, 2020ApJ...901..108V, 2021MNRAS.503.3552A, 2021ApJ...923..201E, 2021ApJ...914...80P, 2021MNRAS.502.3066S, 2021ApJ...914..140P, 2022ApJ...924...38K, 2022MNRAS.516.1752M, 2022MNRAS.514.3941N, 2022PhRvD.105f3018P, 2022PhRvD.105j3008R, 2023MNRAS.520.5622B, 2023PhRvD.108j3036D, 2023PhRvL.131s1201J, 2023PhRvD.107j3025K, 2023PhRvD.107d3008M, 2023ApJ...959...21P, 2023PhRvD.107j3015V, 2023PhRvD.107l3005A, 2023PhRvD.107h3029B, 2023PhRvD.107h3017L, 2024PhRvD.110h3006M, 2025arXiv250306406M}. 
    However, a growing body of work highlights the low-frequency $\lessapprox$ 250 Hertz band of emission, as well as  methods and prospects for its detection
    \cite{1978ApJ...223.1037E, 1978Natur.274..565T, 1996PhRvL..76..352B, 2004ApJ...603..221M, 2009ApJ...697L.133K, 2009ApJ...707.1173M, 2010CQGra..27s4005Y, 2011ApJ...743...30T, 2011ApJ...736..124K, 2012A&A...537A..63M, 2013ApJ...766...43M, 2015PhRvD..92h4040Y, 2018ApJ...861...10M, 2018ApJ...861...10M, 2019MNRAS.487.1178P, 2019ApJ...876L...9R, 2020MNRAS.494.4665P, 2022PhRvD.105j3008R, 2022MNRAS.510.5535J, 2022MNRAS.514.3941N, 2022PhRvD.106d3020M, 2023PhRvD.107d3008M, 2023MNRAS.522.6070P, 2023ApJ...959...21P, 2023PhRvD.107j3015V, 2024PhRvL.133w1401R, 2024ApJ...975...12C, 2021arXiv210505862M, 2024arXiv240513211G}.
    In this investigation, we focus on the permanent deformation of space-time, known as linear gravitational wave memory, the $0$-Hz limit of the signal. The $0$-Hz signal is undetectable in current ground-based detectors and will be undetectable in future, proposed ground-based detectors \cite{2023ApJS..267...29A, 2019BAAS...51g..77T, 2019BAAS...51g..35R, 2020JCAP...03..050M}, but should be detectable in future, space-based detectors, which will move on geodesics. However, while the memory {\em per se} cannot be detected currently or through ground-based detectors, \citet{2024PhRvL.133w1401R} have proposed that the very low-frequency gravitational wave emission, specifically below 50 Hertz, leading up to the final gravitational wave memory---i.e., the ramp up to, and a signature of, the memory---is detectable. This ramp up results from the slowly varying quadrupole moment associated with the nonspherical explosion of the massive star. Here, we further investigate the detectability of low-frequency gravitational wave emission in current ground-based gravitational wave detectors and future ground- and space-based detectors.

    In Section \ref{sec:LFGW}, we discuss the low-frequency gravitational waves from both the evolution of the stellar core fluid (\ref{sec:LFGW-Fluid}) and the anisotropy of the neutrino radiation field (\ref{sec:LFGW-Neutrino}). In Section \ref{sec:LFGWA}, we discuss 
    (i) different analysis and detection techniques, including matched filtering (\ref{sec:LFGWA-Logistic}), 
    (ii) the reconstruction of gravitational waves, sourced from both the fluid motions and the neutrino emission, in real interferometric noise (\ref{sec:LFGWA-cWB}), and 
    (iii)detection prospects for a limited selection of future detectors (\ref{sec:LFGWA-Prospects}). Finally, we discuss our results and conclude in Section \ref{sec:Conclusion}. 
    \begin{figure*}
        \centering
        \includegraphics[width=0.9\linewidth]{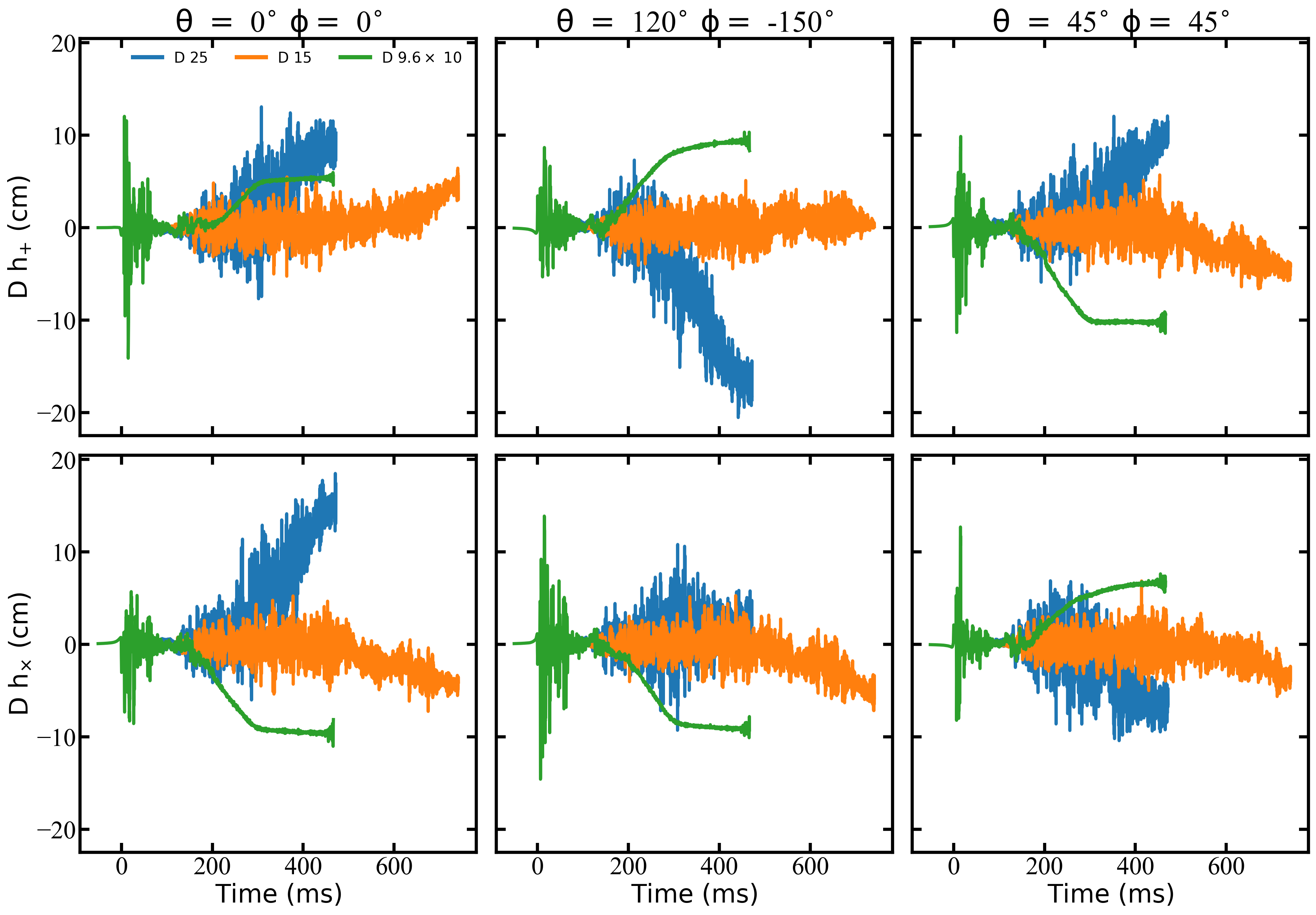}
        \caption{Gravitational waves sourced from the fluid motions in our core-collapse supernova models. The top row shows the plus polarization at the source. The bottom row shows the cross  polarization at the source.  Each column shows the different signals at a  specific observer orientation with respect to the source frame. Note the significantly lower amplitude signal from the D9.6-3D model, which indicates a more spherical explosion.}
        \label{fig:GWs-Flow}
    \end{figure*}
\section{Low-Frequency Gravitational Waves}
\label{sec:LFGW}

    In this section, we present gravitational wave emission predictions from three state-of-the-art three-dimensional models produced using the \textsc{Chimera} core-collapse supernova code \cite{2020ApJS..248...11B, 2023PhRvD.107d3008M}, with a focus on the low-frequency portion of the signals. We present predictions for the low-frequency emission resulting from the nonspherical and time-dependent (i) ejection of stellar material and (ii) emission of neutrinos. The models used here, from the \textsc{Chimera} D-series of simulations and associated with the most recent release of gravitational wave data produced by the \textsc{Chimera} code, are initiated from three different progenitors: $\mathrm{9.6 \  M_{\odot}}$ (D9.6-3D) \cite{2010ApJ...724..341H}, $\mathrm{15 \  M_{\odot}}$ (D15-3D) \cite{2007PhR...442..269W}, and $\mathrm{25 \ M_{\odot}}$ (D25-3D) \cite{2010ApJ...724..341H}.

    All three progenitors are nonrotating. The D9.6-3D and D25-3D progenitors have zero metallicity, and the D15-3D progenitor has Solar metallicity. For D9.6-3D, Figure 16 of \citet{2023PhRvD.107d3008M} shows that the energy carried away in the form of gravitational waves is no longer evolving as a function of time at the end of the run \cite{2021ApJ...921..113S}.

    This is not the case for the D15-3D and D25-3D models, which are still evolving. Although D15-3D and D25-3D are still evolving, we expect the strain---specifically, the memory---to saturate eventually, so we use the incomplete explosion to extrapolate the strain for the detection study below.
    The strain will either saturate to some non-zero value or will return to zero, depending on the dynamics of the event and the sources of gravitational radiation included. For the gravitational waves sourced from the fluid, as the explosion becomes spherical and the densities fall off, the gravitational wave strain should return to zero. However, as shown in Eq. \ref{eq:Nu2GW}, even if the neutrino luminosity becomes spherical at some point in time and the anisotropy goes to zero, the strain sourced from neutrino emission is additive; therefore, the strain will remain constant at the value it had prior to the onset of spherical symmetry. In the analysis presented here, we assume that the fluid contribution will remain constant on time scales up to shock breakout and that the neutrino contribution will remain constant after the end of the simulation. The results from \citet{2012A&A...537A..63M} and \citet{2022PhRvD.105j3008R} on time scales associated with shock break out demonstrate that this is a reasonable assumption. 

    \subsection{Gravitational Waves from Fluid Flow}
    \label{sec:LFGW-Fluid}
        The gravitational waves produced by the motion of the fluid in all three models have already been investigated in \citet{2023PhRvD.107d3008M} and \citet{2025arXiv250306406M}, but this was for the polar and equatorial observer orientations only.

        Here we document a subset of the gravitational waveforms from 2664 different observer orientations, equivalent to 5-degree resolution in both the azimuthal and polar angles. This angular decomposition is done to facilitate the combination of gravitational waves sourced from the fluid motion and the neutrino anisotropy, the latter of which is described in Section \ref{sec:LFGW-Neutrino}.
        [For interested parties, the gravitational waves sourced from the fluid for all observer orientations are available at \href{https://doi.ccs.ornl.gov/dataset/847fc720-6ff7-50eb-a747-12fbb23038db}{Constellation: Chimera D-Series Gravitational Wave Emission Sourced from Neutrino Anisotropy}]. 
        In Fig. \ref{fig:GWs-Flow} we present the plus and cross gravitational wave strains sourced from the fluid only for three observer orientations for all three models. Here we can see the variation in the low-frequency portion of the strains and how they are dependent on the observer location, and how the relative scale of each signal is similar regardless of observer orientation. The extraction of these waveforms is detailed in \citet{2023PhRvD.107d3008M}, but the final expression comes from the quadrupole moment of the transverse-traceless gravitational wave strain \cite{2006RPPh...69..971K}, 
        \begin{equation*}
            \begin{split}
                h_{+} & = \frac{h_{\theta \theta}^{TT}}{r^{2}}, \\
                h_{\times} & = \frac{h_{\theta \phi}^{TT}}{r^{2} \sin{\theta}}, \ \\
                & \mathrm{where} \\
                h_{i j}^{TT} & = \frac{G}{c^{4}} \frac{1}{r} \sum_{m = - 2}^{+2} \frac{d^{2} I_{2 m}}{d t^{2}} \bigg( t - \frac{r}{c} \bigg) f_{i j}^{2m}.
            \end{split}
        \end{equation*}
        Here we make use of the spherical harmonic expansion of the second time derivative of the mass quadrupole moment, $I$, where $f^{2m}_{ij}$ are the tensor spherical harmonics and $i$ and $j$ run over $r$, $\theta$, and $\phi$.

    \subsection{Gravitational Waves from Anisotropic Neutrino Emission}
    \label{sec:LFGW-Neutrino}
        As first proposed by \citet{1978ApJ...223.1037E}, the aspherical emission of neutrinos also produces gravitational radiation. 
        In order to extract the gravitational waves sourced from the anisotropic neutrino emission we begin with the methodology proposed by \citet{1978ApJ...223.1037E}. Starting from the Einstein field equations as presented in \citet{1973grav.book.....M},
         \begin{equation}
             G_{\mu \nu} = 8 \pi T_{\mu \nu},
         \end{equation}
        we introduce a linear perturbation to the metric,
        \begin{equation}
            g_{\mu \nu}= \eta_{\mu \nu} + h_{\mu \nu} +  \mathcal{O}(h^{2}).
        \end{equation}
        Expanding the Einstein tensor and keeping only the terms linear in $h$, we find
        \begin{equation}
            \begin{split}
                G_{\mu \nu} & = R_{\mu \nu} - \frac{1}{2} g_{\mu \nu} R \\
                & = \frac{1}{2} (h_{\nu \alpha \ \ , \mu}^{ \ \ \ , \alpha} - \Box h_{\mu \nu} - h^{\alpha}_{ \ \ \alpha, \mu \nu} + h_{\mu \alpha \ \ , \nu}^{ \ \ \ , \alpha}) \\
                & - \frac{1}{2} \eta_{\mu \nu} (- \Box h + h_{\alpha \beta}^{ \ \ \ , \alpha \beta}).
            \end{split}
        \end{equation}
        Finally, we introduce a new gauge,
        \begin{equation}
            \begin{split}
                \bar{h}_{\mu \nu} & = h_{\mu \nu} - \frac{1}{2} \eta_{\mu \nu} h, \\
                h_{\mu \nu} & = \bar{h}_{\mu \nu} + \frac{1}{2} \eta_{\mu \nu} h = \bar{h}_{\mu \nu} + \frac{1}{2} \eta_{\mu \nu} \bar{h}, \\
            \end{split}
        \end{equation}
        which allows us to rewrite the Einstein tensor and subsequent Einstein equations as
        \begin{equation}
            G_{\mu \nu} = - \frac{1}{2} \Box \bar{h}_{\mu \nu} = 8 \pi T_{\mu \nu}.
        \end{equation}
        From here we finally arrive at the wave equation for $\bar{h}$,
        \begin{equation}
            \Box \bar{h}_{\mu \nu} = -16 \pi T_{\mu \nu}.
        \end{equation}
        This wave equation can be solved with a retarded Green's function of the form,
        \begin{equation}
            \begin{split}
                \Box G(x - x') & = \delta^{4}(x - x') \\
                & \mathrm{ \ and} \\
                G(x - x') & = - \frac{1}{4 \pi |x - x'|} \delta(x_{r}^{0} - x'^{0}), \\
            \end{split}
        \end{equation}
        where $x_{r}^{0} = t - |\vec{x} - \vec{x}'|$.
        This allows us to write the solution to the wave equation as
        \begin{equation}
            \label{eq:pre_strain}
            \begin{split}
                \bar{h}_{\mu \nu}(x) & = -16 \pi \int d^{4} x' G(x - x') T_{\mu \nu}(x') \\
                & = 4 \int d^{3} x' \frac{1}{|\vec{x} - \vec{x}'|}T_{\mu \nu}(x^{0}_{r}, \vec{x}').
            \end{split}
        \end{equation}
        From here we motivate a form of the stress-energy tensor following \citet{1978ApJ...223.1037E},
        \begin{equation}
        \label{eq:stress}
            T_{i j}(x) = \frac{n_{i} n_{j}}{r^{2}} \int_{- \infty}^{\infty} dt' \sigma(t') f(t',\Omega') \delta(t - t' - r),
        \end{equation}
        where $\hat{n} = \vec{x} / r$, $|\vec{x}|=r$, $f(t',\Omega')$, and $\sigma(t')$ represents, in Eq. (\ref{eq:stress}), matter being continuously released at the speed of light from the point $\vec{x}=\vec{0}$. The functions $\sigma(t')$ and $f(t',\Omega')$ are the rate of energy loss and the angular distribution of the emission at time $t'$, respectively. This form of the stress-energy tensor allows us to write the gravitational wave strain as
        \begin{equation}
            \begin{split}
                \bar{h}_{i j}(x) = & 4 \int_{0}^{\infty} \int_{4 \pi} 
            \int_{0}^{\infty} dr' d\Omega' dt' \\ 
            & \frac{n_{i} n_{j}}{|\vec{x} - \vec{x}'|} \sigma(t') f(t',\Omega') \delta(t - |\vec{x} - \vec{x}'| - t' - r).
            \end{split}
        \end{equation}
        We first consider the integral over $r'$,
        \begin{equation}
            I = \frac{1}{2} \int_{-\infty}^{\infty} dr'\frac{1}{|\vec{x} - \vec{x}'|} \delta(t - |\vec{x} - \vec{x}'| - t' - r),
        \end{equation}
        where we expand $|\vec{x} - \vec{x}'|^{2} = r^{2} + r'^{2} - 2 r r' \cos{\theta}$ and use the Dirac delta property $\delta(g(x)) = \frac{\delta(x - x_{0})}{|g'(x_{0})|}$ to evaluate the integral, to obtain
        \begin{equation}
            I  = \frac{1}{2} \frac{1}{|g'(r'_{0})|} \frac{1}{|\vec{x} - \vec{x}'_{0}|},
        \end{equation}
        where 
        \begin{equation}
        \label{eq:integral_delta_sol}
            \begin{split}
                g'(r') & = - \frac{r' - r \cos{\theta} + |\vec{x} - \vec{x}'|}{|\vec{x} - \vec{x}'|}, \\
                r'_{0} & = - \frac{r^{2}-\tau^{2}}{2 (\tau - r \cos{\theta})}, \\
                \mathrm{and} & \\
                \tau =&  t - t' \  \mathrm{for} \ \tau \geq r. \\
            \end{split}
        \end{equation}
        From here we calculate $g'(r'_{0})$ and $|\vec{x} - \vec{x}'_{0}|$ and find that
        \begin{equation}
            \label{eq:intermediate_integral}
            I = \frac{1}{2} (r'_{0} - r \cos{\theta} + |\vec{x} - \vec{x}'_{0}|)^{-1}.
        \end{equation}
        Focusing on just the denominator alone, and using Eqs. \ref{eq:integral_delta_sol}, we find,
        
        \begin{equation}
            \begin{split}
                & r'_{0} - r \cos{\theta} + |\vec{x} - \vec{x}'_{0}| = \\
                & \frac{1}{2(\tau - r \cos{\theta})} (-r^{2} + \tau^{2} - 2 (\tau -r \cos{\theta}) r \cos{\theta} + \\
                & [(2 (\tau- r \cos{\theta}))^{2} r^{2} + (r^{2} - \tau^{2})^{2} + \\
                & 2 r (2 (\tau - r \cos{\theta})) (r^{2} -\tau^{2}) \cos{\theta}]^{1/2}). \\
            \end{split}
        \end{equation}
        We can simplify the expression in the square-root to $(r^{2} + \tau^{2} - 2 r \tau \cos{\theta})^{2}$, which allows us to write,
        \begin{equation}
            r'_{0} - r \cos{\theta} + |\vec{x} - \vec{x}'_{0}| = \frac{1}{2(\tau - r \cos{\theta})} (\tau - r \cos{\theta})^{2}.
        \end{equation}
        Using this in Eq. (\ref{eq:intermediate_integral}) gives
        \begin{equation}
            \label{eq:intermediate_sol}
            I = \frac{1}{2} \frac{1}{r'_{0} - r \cos{\theta} + |\vec{x} - \vec{x}'_{0}|} = \frac{1}{(t - t' - r \cos{\theta})}.
        \end{equation}
        Finally using Eq. (\ref{eq:intermediate_sol}) in Eq. (\ref{eq:pre_strain}), we obtain 
        \begin{equation}
            \bar{h}_{i j}(x) = 4 \int_{- \infty}^{t - r} \int_{4 \pi} d\Omega' dt' \frac{n_{i} n_{j}}{t - t' - r \cos{\theta}} \sigma(t') f(t',\Omega') .
        \end{equation}
        With this general form of the strain, we project into the transverse-traceless gauge, which only affects the directional vector, $\hat{n}$, resulting in Equation 16 of \citet{1978ApJ...223.1037E}:
        \begin{equation}
            \bar{h}_{i j}^{TT}(x) = 4 \int_{- \infty}^{t - r} \int_{4 \pi} d\Omega' dt' \frac{(n_{i} n_{j})^{TT}}{t - t' - r \cos{\theta}} \sigma(t') f(t',\Omega').
        \end{equation}

\begin{figure}
            \centering
            \includegraphics[width=0.99\linewidth]{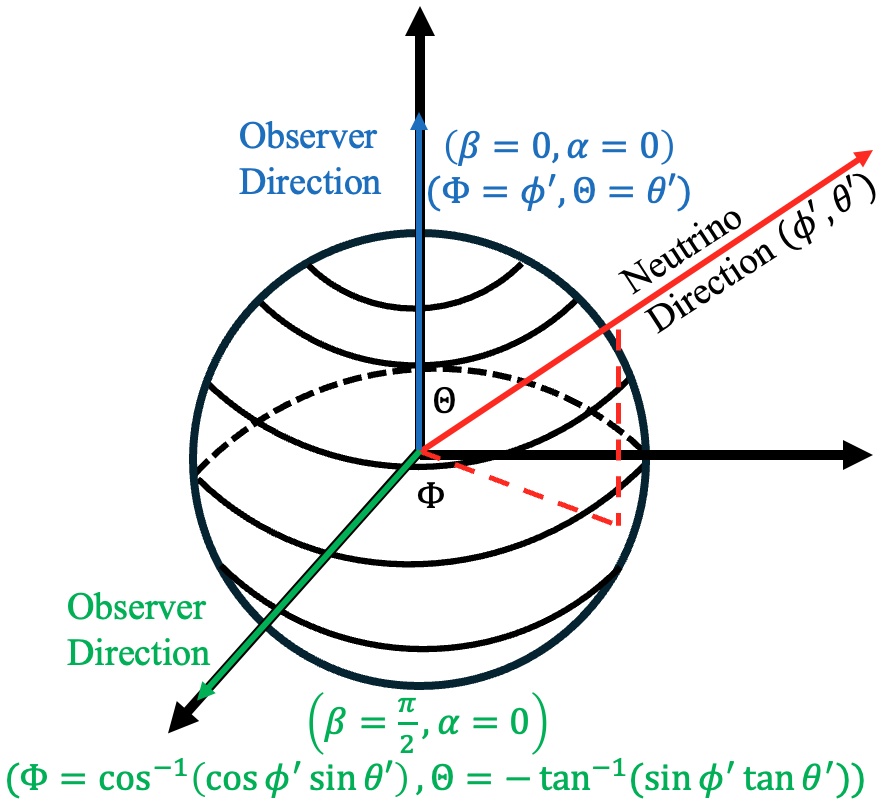}
            \caption{The spherical polar frame of the source is shown in black, where the radiation in a solid angle $(\phi,\theta)$ being emitted from the source is shown in red. We superimpose two example observer frames, given by $(\beta,\alpha)$. The frame delineated in blue corresponds to the observer frame assuming the observer is along the source's z-axis. The frame delineated in green corresponds to the observer frame assuming the observer is along the source's x-axis.}
            \label{fig:relative_obs_source}
        \end{figure}

        From this point we follow \citet{1997A&A...317..140M} and calculate the transverse-traceless portion of $\bar{h}_{i j}$, which only involves projecting $n_{i} n_{j}$. We begin by assuming our observer is located at a distance $R$ along the z-axis, which results in their unit vector $\hat{\tilde{n}} = (0,0,1)$. We follow \citet{1973grav.book.....M} and define our transverse-traceless projection operator,
        \begin{equation}
            \begin{split}
                \Lambda_{i j}^{k l} & = P_{i}^{k} P_{j}^{l} - \frac{1}{2} P_{i j} P^{k l}, \\
                & \ \mathrm{with} \\  
                P^{i}_{j} & = \delta^{i}_{j} - \tilde{n}^{i} \tilde{n}_{j}.
            \end{split}
        \end{equation}
        This projection, $(n_{i} n_{j})^{TT} = \Lambda_{i j}^{k l} n_{k} n_{l}$, results 
        in only $n_{x} n_{x}$, $n_{y} n_{y}$ and $n_{x} n_{y} = n_{y} n_{x}$ surviving. As this vector represents the direction of material leaving the system in the observer frame, we define $\hat{n} = (\cos{\phi} \sin{\theta}, \sin{\phi} \sin{\theta}, \cos{\theta})$, so that
        \begin{equation}
            \begin{split}
                (n_{x} n_{x})^{TT} & = \frac{1}{2} (\cos^{2}{\phi} \sin^{2}{\theta} - \sin^{2}{\phi} \sin^{2}{\theta}) \\
                & = \frac{1}{2} (1 - \cos^{2}{\theta}) (2 \cos^{2}{\phi} - 1) \\
                & = \frac{1}{2} (1 - \cos{\theta}) (1 + \cos{\theta}) \cos{2 \phi}\\
                (n_{y} n_{y})^{TT} & = - (n_{x} n_{x})^{TT} \\
                (n_{x} n_{y})^{TT} & = \cos{\phi} \sin{\theta} \sin{\phi} \sin{\theta} \\
                & = (1 - \cos^{2}{\theta}) (\cos{\phi} \sin{\theta}) \\
                & = \frac{1}{2} (1 - \cos{\theta}) (1 + \cos{\theta}) \sin{2 \phi}. \\
            \end{split}
        \end{equation}
        With these projections in hand, we arrive at the gravitational wave strain sourced from continuous emission of neutrinos from the core of our CCSN:
        \begin{equation}
        \label{eq:GWfromNu}
            \begin{split}
                h_{+}(x) & = 2 \int_{- \infty}^{t - r} \int_{4 \pi} d\Omega' dt' \frac{(1 - \cos^{2}{\theta}) \cos{2 \phi}}{t - t' - r \cos{\theta}} \sigma(t'), f(t',\Omega') \\
                h_{\times}(x) & = 2 \int_{- \infty}^{t - r} \int_{4 \pi} d\Omega' dt' \frac{(1 - \cos^{2}{\theta}) \sin{2 \phi}}{t - t' - r \cos{\theta}} \sigma(t') f(t',\Omega').
            \end{split}
        \end{equation}
        It is important to note that $\theta$ and $\phi$ are relative to the observer, not the source, while the radiation's angular distribution is defined in the source frame. Fig. \ref{fig:relative_obs_source} shows the relationship between the observer frame and the source frame. 
        By assuming that $t' - t = r$---i.e., by assuming that the gravitational waves are produced only by the neutrino pulse---we simplify the denominator of Eq. \ref{eq:GWfromNu}: $t - t' - r\cos{\Theta} = r (1 - \cos{\Theta})$.
        By utilizing the ray-by-ray nature of the \textsc{Chimera} neutrino transport we can directly compute the directional loss of energy from the radiation fields
        \begin{equation*}
            \frac{d L}{d \Omega'} (t', \Omega') \sim \sigma(t') f(t', \Omega').
        \end{equation*}

        Within \textsc{Chimera} neutrinos are treated classically ---i.e. as massless fermions - and are transported in the ray-by-ray approximation with flux-limited diffusion to close the moment equations. Distributions for electron, anti-electron, heavy (muon + tau) and anti-heavy (anti-muon + anti-tau) neutrinos are evolved. The quantity $\frac{d L}{d \Omega'} (\Omega', t')$ is the total neutrino luminosity at a time $t'$ in a solid angle $d\Omega'$ in a direction $(\theta', \phi')$. In \textsc{Chimera} the differential neutrino luminosity, denoted by $\frac{d L_{E}^{\nu_{i}}}{d\Omega}$ for each neutrino species $\nu_{i}$, is computed using
        \begin{figure}
            \centering
            \includegraphics[width=0.9\linewidth]{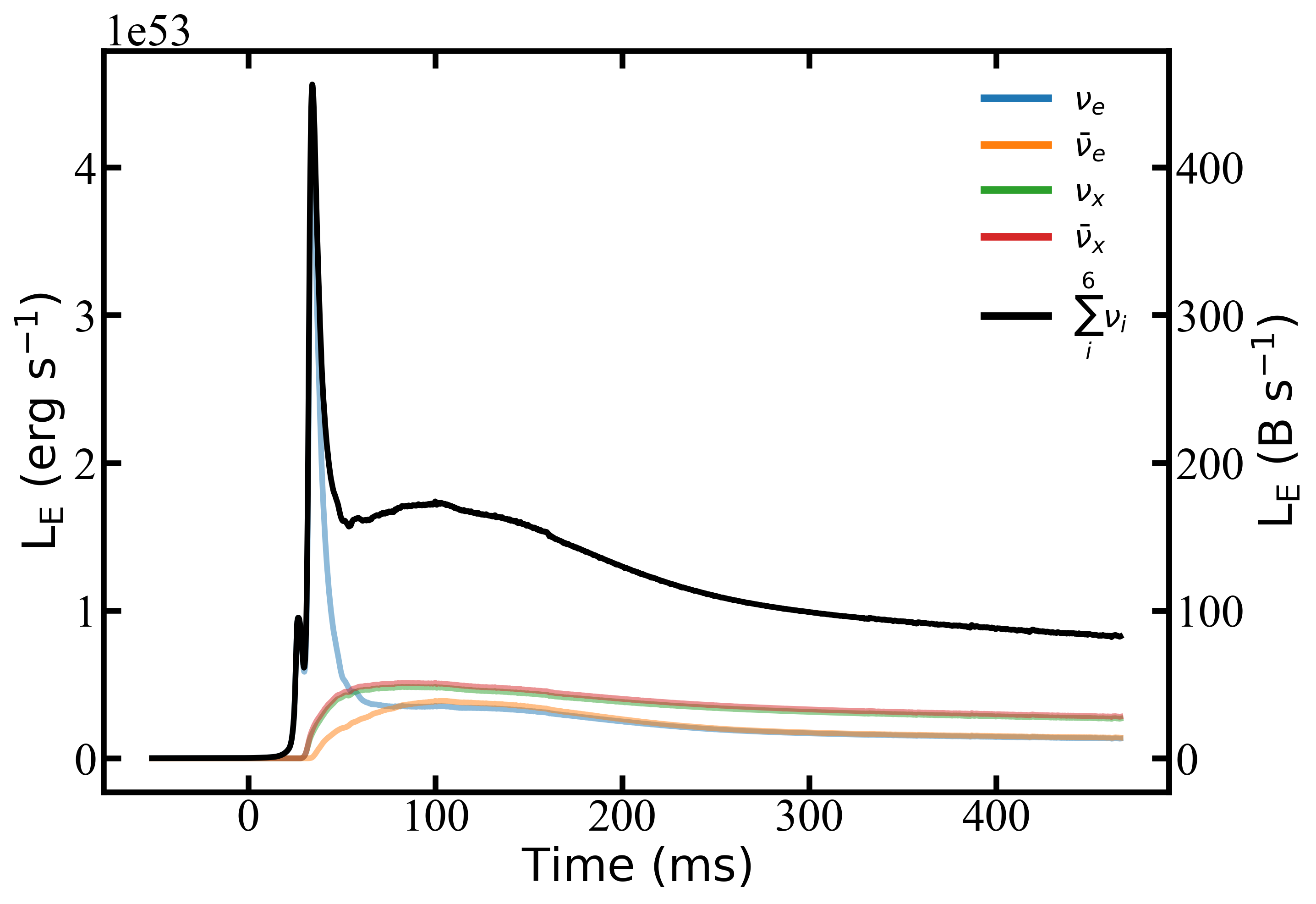}
            \includegraphics[width=0.9\linewidth]{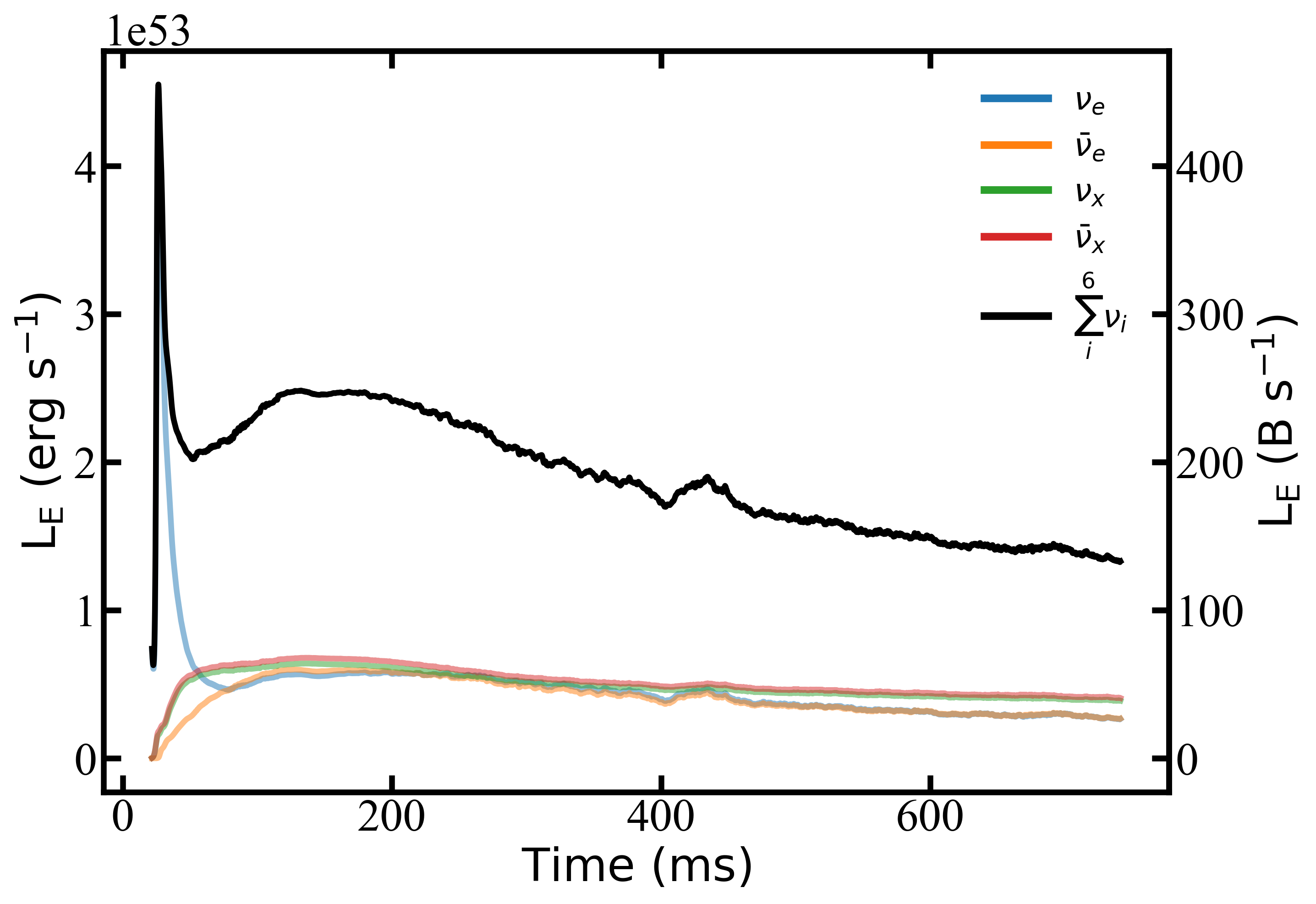}
            \includegraphics[width=0.9\linewidth]{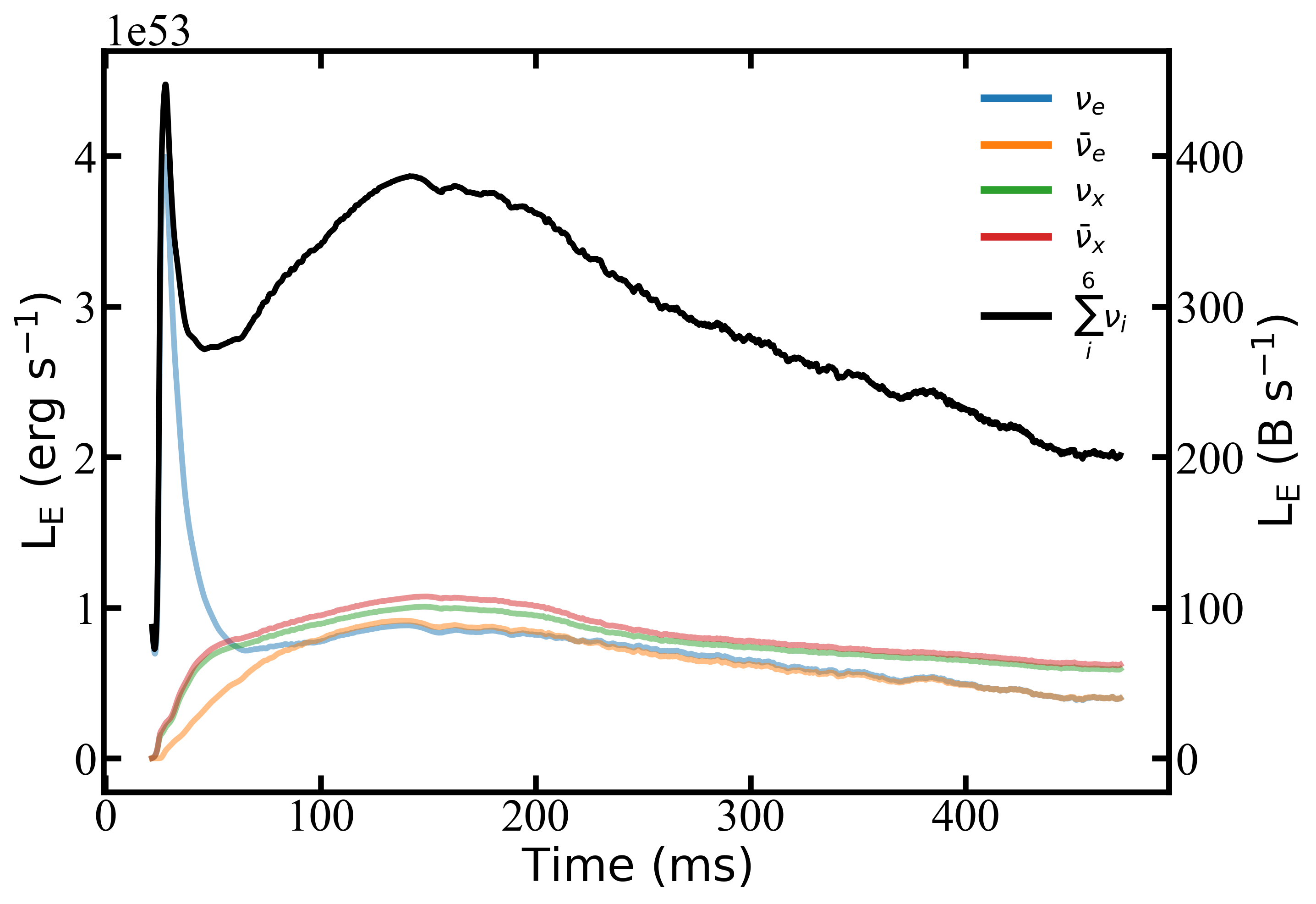}
            \caption{Energy luminosity of the neutrinos from D9.6-3D (top), D15-3D (middle), and D25-3D (bottom). The luminosity for each evolved species separately is delineated by color, and the total neutrino luminosity is delineated in black.}
            \label{fig:lum}
        \end{figure}
        
        \begin{equation}
         \frac{d L_{E}^{\nu_{i}}}{d \Omega}(t,\Omega) = \frac{4 \pi r^{2} c}{(h c)^{3}} \frac{1}{\alpha^{4}} \int dE \psi^{(1)}(\phi, \theta, r, \nu_{i}, E) E^{3},
        \end{equation}
        It is extracted at $r=500$ km from the first moment of the neutrino distribution, $\psi^{(1)}$, computed in the Eulerian-Lab frame of reference. We also correct for the gravitational energy shift with the lapse, $\alpha$, and introduce the global neutrino energy luminosity passing though a spherical surface at the radius of extraction as
        \begin{equation}
            L^{\nu}_{E}(t) = \int_{4 \pi} d \Omega' \frac{d L^{\nu}} {d\Omega'}(\Omega', t').
        \end{equation}
        Fig. \ref{fig:lum} shows this angle integrated neutrino luminosity for each neutrino species and for each model.
        
\begin{figure}
                \centering
                \includegraphics[width=0.9\linewidth]{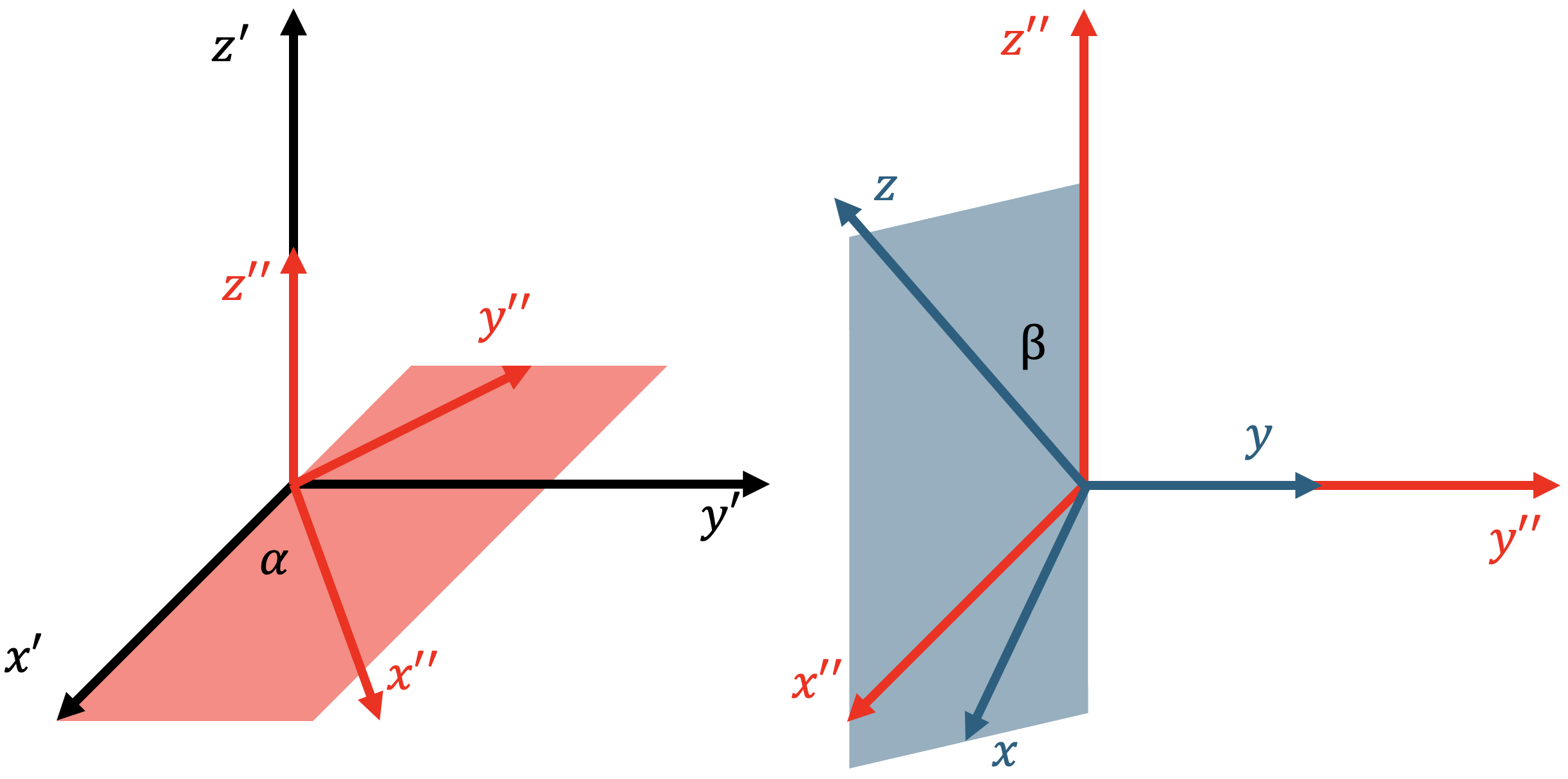}
                \caption{Observer--source relative orientation, reproduced from \citet{2012A&A...537A..63M}.}
                \label{fig:angles}
            \end{figure}

        The plus and cross polarizations of the gravitational wave strain along the z-axis of the observer are concisely expressed, as in \citet{1997A&A...317..140M} and \citet{2012A&A...537A..63M}, as
                \begin{equation}
                \label{eq:GWPlusNu}
                    h_{+}(t) = \frac{2}{r} \int_{0}^{t} \int_{4 \pi} dt' d\Omega' (1 + \cos{(\theta)})\cos{(2 \phi)} \frac{d L} {d\Omega'}(\Omega', t'),
                \end{equation}
                \begin{equation}
                \label{eq:GWCrossNu}
                    h_{\times}(t) = \frac{2}{r} \int_{0}^{t} \int_{4 \pi} dt' d\Omega' (1 + \cos{(\theta)})\sin{(2 \phi)} \frac{d L} {d\Omega'}(\Omega', t'),
                \end{equation}
        Following \citet{2012A&A...537A..63M} we now introduce two viewing angles, $\alpha \in [-\pi, \pi]$, viewed as a rotation in the x-y plane of the source, and $\beta \in [0, \pi]$, viewed as a rotation in the new x-z plane (see Fig. \ref{fig:angles}). We begin with the $\alpha$ rotation (corresponding to the left panel of Fig. \ref{fig:angles}),
        \begin{equation}
            \begin{split}
                x'' & = x' \cos{\alpha} + y' \sin{\alpha}, \\
                y'' & = -x' \sin{\alpha} + y' \cos{\alpha}, \ \\
                & \mathrm{and}\\
                z'' & = z'. \\
            \end{split}
        \end{equation}
        Next we rotate through an angle $\beta$ in the new $x''$-$z''$ plane (corresponding to the right panel of Fig. \ref{fig:angles}),
        \begin{equation}
            \begin{split}
                x & = x'' \cos{\beta} - z'' \sin{\beta}, \\
                y & = y'', \ \\
                & \mathrm{and}\\
                z & = x'' \sin{\beta} + z'' \cos{\beta}. \\
            \end{split}
        \end{equation}
        We now relate $x$, $y$, and $z$ to $\alpha$, $\beta$, $\theta'$, $\phi'$, $\theta$, and $\phi$, starting with the spherical polar coordinates of the observer,
        \begin{equation}
        \label{eq:obs_sph-polar}
            \begin{split}
                \frac{x}{r} & = \cos{\phi} \sin{\theta} \\
                            & = \cos{\beta} \sin{\theta'} [\cos{\alpha} \cos{\phi'} + \sin{\alpha} \sin{\phi'}] - \sin{\beta} \cos{\theta'}, \\ 
                \frac{y}{r} & = \sin{\phi} \sin{\theta}, \\
                            & = \sin{\theta'} [\cos{\alpha} \sin{\phi'} - \sin{\alpha} \cos{\phi'}], \\ 
                            & \mathrm{and} \\
                \frac{z}{r} & = \cos{\theta}, \\
                            & = \sin{\beta} \sin{\theta'} [\cos{\alpha} \cos{\phi'} + \sin{\alpha} \sin{\phi'}] + \cos{\beta} \cos{\theta'}. \\ 
            \end{split}
        \end{equation}
        In order to calculate the strains from our new observer along $(\alpha, \beta)$ we need to calculate $\cos{\theta}$, $\cos{2 \phi}$, and $\sin{2\phi}$,
        \begin{equation}
        \label{eq:obs_source}
            \begin{split}
                \cos{\theta} &  = \frac{z}{r}, \\
                \cos{2 \phi} & = \frac{2 x y}{x^{2} + y^{2}}, \ \\
                & \mathrm{and} \\
                \sin{2 \phi} & = \frac{x^{2} - y^{2}}{x^{2} + y^{2}}. \\
            \end{split}
        \end{equation}
        Eqs. \ref{eq:obs_sph-polar} into Eqs. \ref{eq:obs_source} yield
        \begin{equation}
            \begin{split}
                & (1 + \frac{z}{r})\frac{r^{2}}{x^{2} + y^{2}} = \\
                & r^{2} \frac{1 + \cos{\beta} \cos{\theta'} + \sin{\beta} \sin{\theta'} [\cos{\alpha} \cos{\phi'} + \sin{\alpha} \sin{\phi'}]}{[\sin{\beta} \cos{\theta'} - \cos{(\alpha -\phi')} \cos{\beta} \sin{\theta'}]^{2}+ \sin^{2}{(\alpha -\phi')} \sin^2{\theta'}},  \\
                & \frac{2 x y}{r^{2}} = \\
                & \frac{1}{r^{2}}[\sin{\beta} \cos{\theta'} - \cos{(\alpha - \phi')} \cos{\beta} \sin{\theta'}]^{2} - \sin^{2}{(\alpha - \phi')} \sin^{2}{\theta'}, \ \\
                & \mathrm{and} \\
                & \frac{x^{2} - y^{2}}{r^{2}} = \\
                & \frac{1}{r^{2}} 2 \sin{(\alpha - \phi')} \sin{\theta'} [\sin{\beta} \cos{\theta'} - \cos{(\alpha - \phi')} \cos{\beta} \sin{\theta'}]. \\
            \end{split}
        \end{equation}
        We can now define the full angular weights,
        \begin{equation}
        \label{eq:angular_weights}
            \begin{split}
                W_{+}(\alpha, \beta, \theta', \phi') & =  (1 + \frac{z}{r}) \frac{x^{2} - y^{2}}{x^{2} + y^{2}} \ \\
                & \mathrm{and} \\
                W_{\times}(\alpha, \beta, \theta', \phi') & =  (1 + \frac{z}{r}) \frac{2 x y}{x^{2} + y^{2}}. \\
            \end{split}
        \end{equation}
        Expanding Eqs. \ref{eq:angular_weights} in $\alpha$, $\beta$, $\theta'$, and $\phi'$ we arrive at the weights presented in \citet{2012A&A...537A..63M}. Inserting the weights from Eqs. \ref{eq:angular_weights} into Eq. \ref{eq:GWCrossNu} we find,
        \begin{equation}
            h^{\nu_{i}}_{+/\times}(t,\alpha, \beta) = \frac{2}{r} \int_{0}^{t} \int_{4 \pi} dt' d\Omega' W_{+/\times}(\alpha, \beta, \Omega') \frac{d L^{\nu_{i}}_{E}} {d\Omega'}(\Omega', t').
        \end{equation}
        Focusing on the $\Omega'$ integral and following \citet{2012A&A...537A..63M} we introduce an anisotropy parameter,
        \begin{equation}
        \label{eq:Anisotropy}
            \alpha^{\nu_{i}}_{+/\times}(t, \alpha, \beta) = \frac{1}{L^{\nu_{i}}_{E}(t)} \int_{4 \pi} d \Omega' W_{+/\times}(\alpha, \beta, \Omega') \frac{d L^{\nu_{i}}_{E}} {d\Omega'}(\Omega', t').
        \end{equation}
        This allows us to write the gravitational wave strain sourced from the anisotropic emission of neutrinos at time $t$ and observer orientation relative to the z-axis of the source $(\alpha, \beta)$ as
        \begin{equation}
        \label{eq:Nu2GW}
            h^{\nu_{i}}_{+/\times}(t, \alpha,\beta) = \frac{2 G}{c^{4} r} \int_{-\infty}^{t - r/c} dt' L^{\nu_{i}}_{E}(t) \alpha^{\nu_{i}}_{+/\times}(t, \alpha, \beta).
        \end{equation}
        
\begin{figure*}
            \centering
            \includegraphics[width=0.9\linewidth]{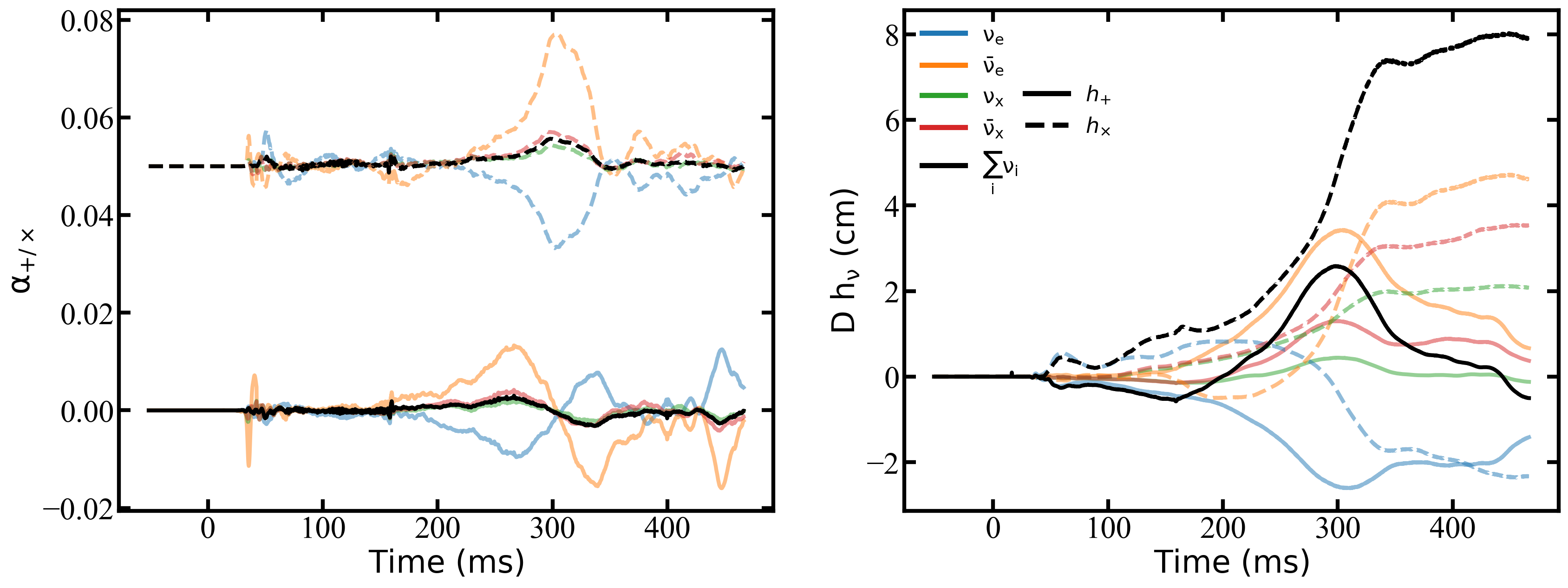}
            \caption{The anisotropy (left) and strain (right) for the plus (solid) and cross (dashed and y-shifted) polarizations for the polar axis and the D9.6-3D model. The contribution of each species independently is delineated by color, and the total is delineated in black.}
            \label{fig:ploar_D9.6}
        \end{figure*}

\begin{figure*}
            \centering
            \includegraphics[width=0.9\linewidth]{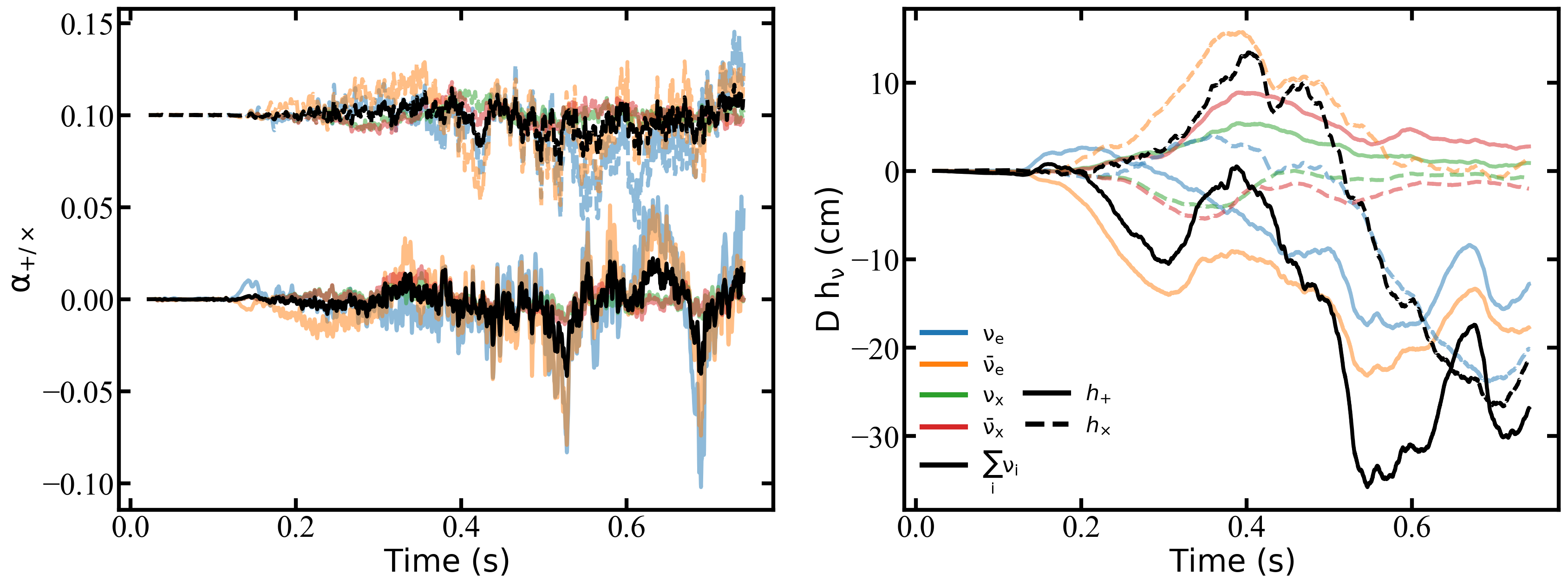}
            \caption{The anisotropy (left) and strain (right) for the plus (solid) and cross (dashed and y-shifted) polarizations for the polar axis and the D15-3D model. The contribution of each species independently is delineated in color, and the total is delineated in black.}
            \label{fig:ploar_D15}
        \end{figure*}

\begin{figure*}
            \centering
            \includegraphics[width=0.9\linewidth]{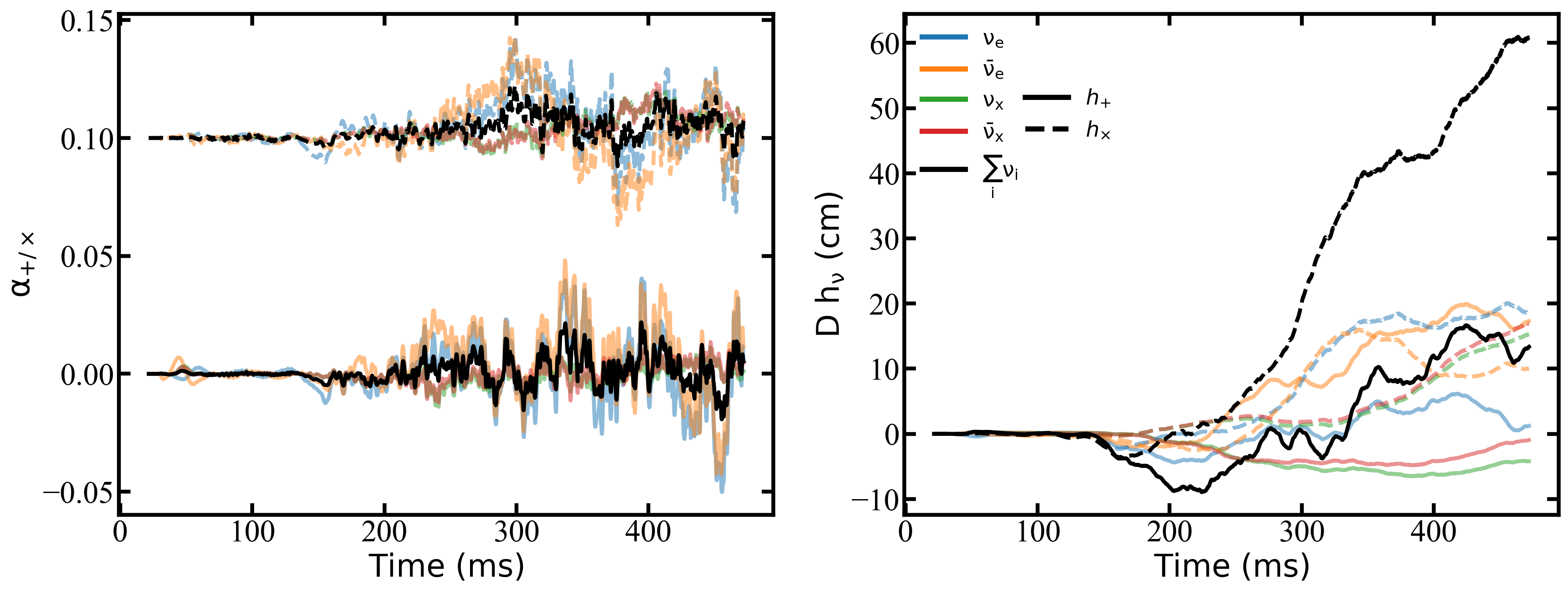}
            \caption{The anisotropy (left) and strain (right) for the plus (solid) and cross (dashed and y-shifted) polarizations for the polar axis and the D25-3D model. The contribution of each species independently is delineated in color, and the total is delineated in black.}
            \label{fig:ploar_D25}
        \end{figure*}

        Having now defined the anisotropy parameter [Eq. (\ref{eq:Anisotropy})] and thus the gravitational wave strain sourced from the anisotropic emission of neutrinos [Eq. (\ref{eq:Nu2GW})] we can investigate their temporal evolution. Figs. \ref{fig:ploar_D9.6}, \ref{fig:ploar_D15}, and \ref{fig:ploar_D25} show the anisotropy (left) and strain (right) for the signal an observer along the polar axis of the source would see for D9.6-3D, D15-3D, and D25-3D, respectively. In all cases, the anisotropy remains below 5\%, indicating a relatively ``spherical'' emission of neutrinos from this point of view, but even this low amount of anisotropy produces strains comparable to the strain sourced from the matter. The anisotropy, and subsequent strain, given for the D9.6-3D model has little high-frequency variation and lower overall amplitude, indicating that the neutrino fields are fairly spherically symmetric over time and that any deviations from spherical symmetry occur on longer time scales. Similar to the gravitational waves sourced from the fluids, this leads to an overall smaller-amplitude strain with no high-frequency components. 
        
\begin{figure}
            \centering
            \includegraphics[width=0.9\linewidth]{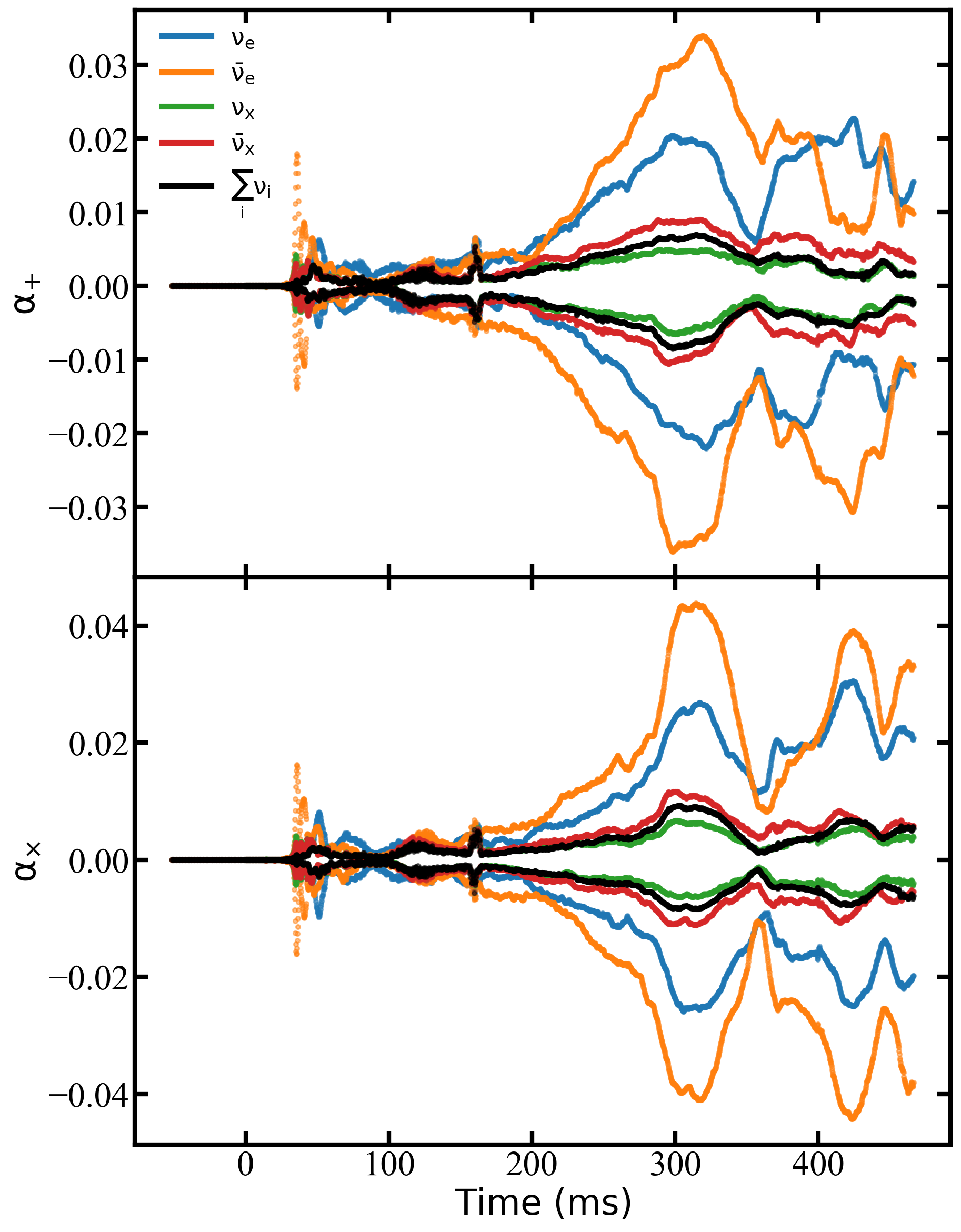}
            \caption{Minimum and maximum anisotropy parameter for D9.6-3D as a function of post--bounce time.}
            \label{fig:min_max_anisotropy_D9.6}
        \end{figure}

\begin{figure}
            \centering
            \includegraphics[width=0.9\linewidth]{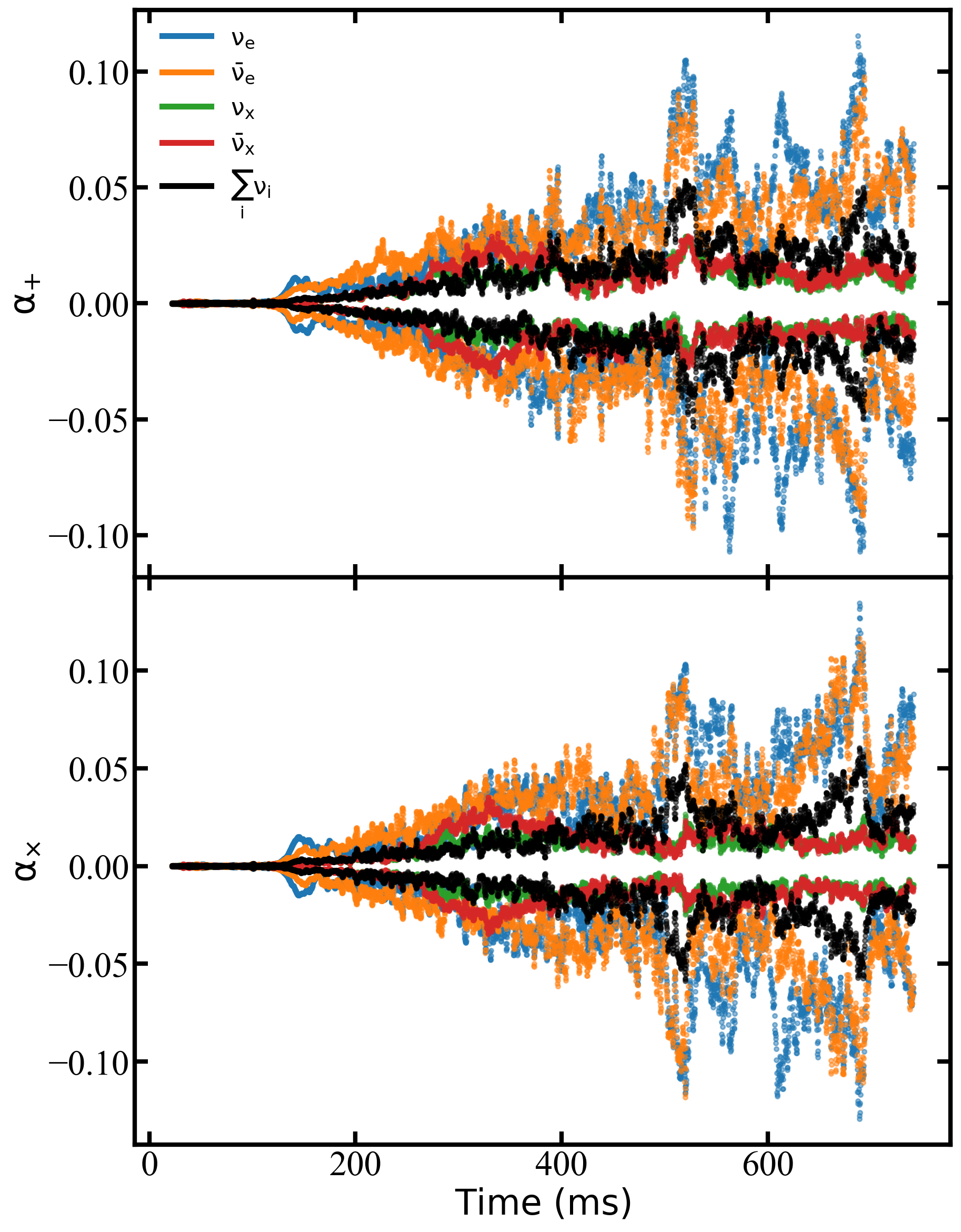}
            \caption{Minimum and maximum anisotropy parameter for D15-3D as a function of post--bounce time.}
            \label{fig:min_max_anisotropy_D15}
        \end{figure}

\begin{figure}
            \centering
            \includegraphics[width=0.9\linewidth]{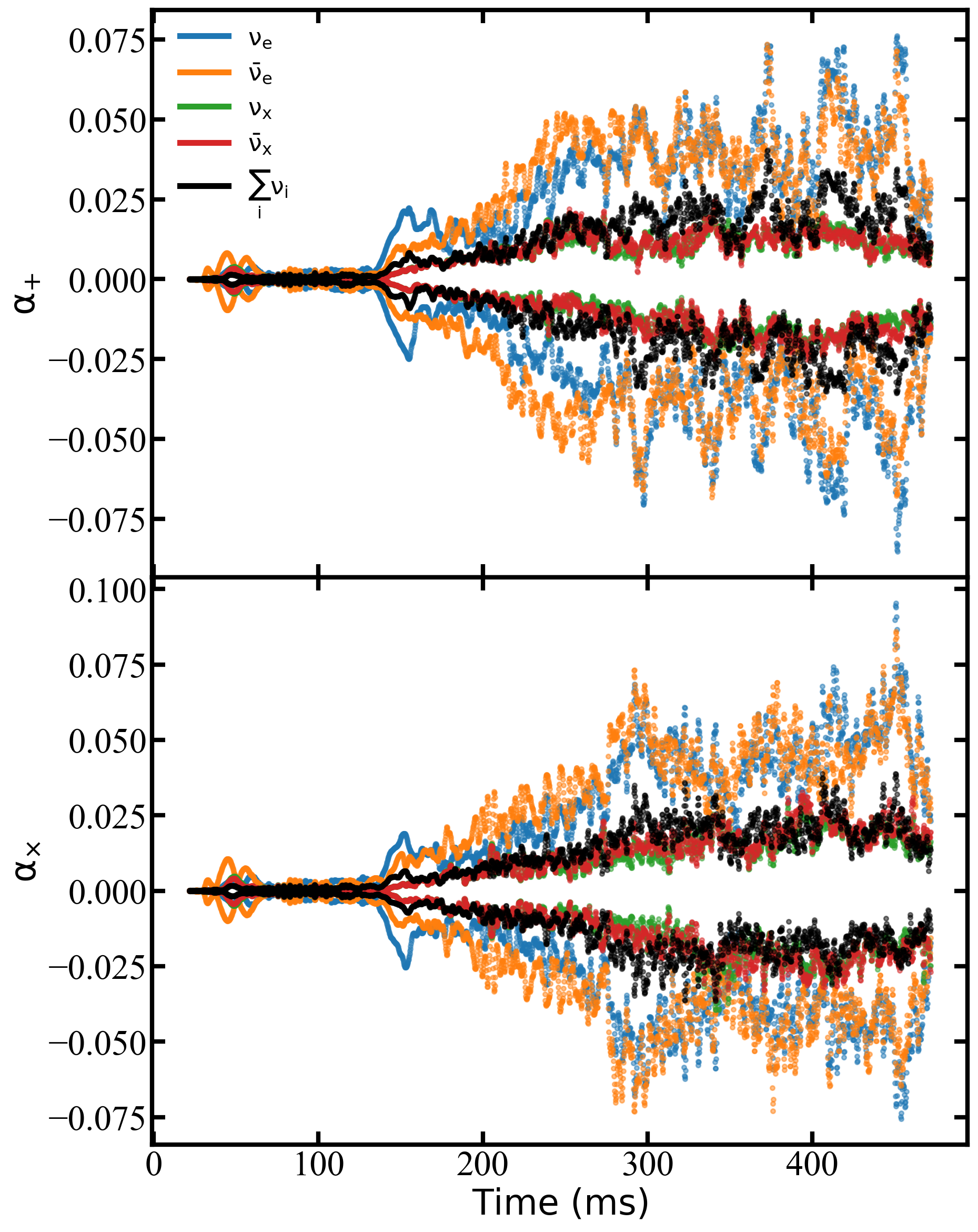}
            \caption{Minimum and maximum anisotropy parameter for D25-3D as a function of post--bounce time.}
            \label{fig:min_max_anisotropy_D25}
        \end{figure}

\begin{figure*}
            \centering
            \includegraphics[width=0.9\linewidth]{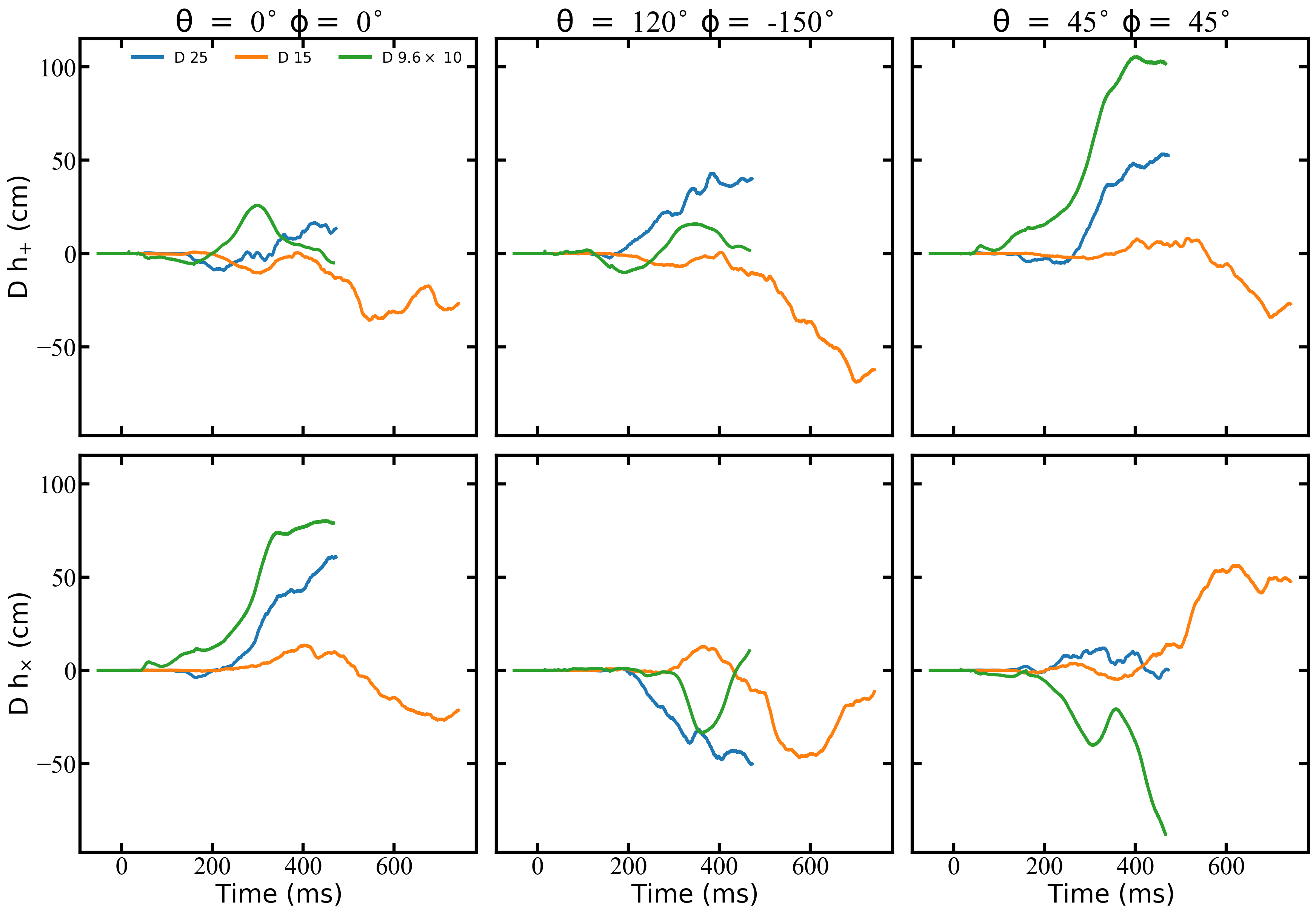}
            \caption{Gravitational waves sourced from the anisotropy of neutrino emission in our core-collapse supernova models. The top row shows the plus polarization at the source. The bottom row shows the cross polarization at the source. Each column shows the different signals at a specific observer orientation with respect to the source frame. Note the significantly lower amplitude signal from the D9.6-3D model, which indicates a more spherical explosion.}
            \label{fig:GWs-Neutrino}
        \end{figure*}

        While the gravitational wave strain from the anisotropic emission of neutrinos that we detect will be from the entire neutrino field, we provide a breakdown of the anisotropy and strain per species evolved by \textsc{Chimera}. In this way, we can see the contribution from each species individually. It is interesting to note that, for the D9.6-3D signals shown in Fig. \ref{fig:ploar_D9.6}, the anisotropy, shown on the left-hand side of the figure, of the electron-neutrino field and the anisotropy of the anti-electron-neutrino field seem to cancel each other out, which leaves the overall anisotropy of the neutrino radiation field dominated by the heavy and anti-heavy neutrino radiation fields. However, as shown on the right-hand side of the figure, all four species contribute to the total gravitational wave strain.
        In the case of D15-3D and D25-3D, no such cancellation is evident in the anisotropy parameters, and, as before, the strains are dependent on all four species. 
        In Figs. \ref{fig:min_max_anisotropy_D9.6}--\ref{fig:min_max_anisotropy_D25}, we provide plots showing the minimum and maximum of the anisotropy parameter, across all observer orientations, for each of the models considered here. Across all observer orientations, the maximum anisotropy remains low, which, as in the polar case described previously, indicates that the emission of neutrinos is fairly spherical regardless of the observer orientation. Again, we see that the overall shape of the total anisotropy is dominated by the heavy and anti-heavy neutrino anisotropies.
        It is also interesting to note that, across all observer orientations and all models investigated here, the gravitational waves from electron neutrinos seem to dominate the high-frequency portion of the anisotropy and subsequent strain. 
        [For interested parties, as in the case of the gravitational waveforms sourced by the fluid flow, we make available \href{https://doi.ccs.ornl.gov/dataset/847fc720-6ff7-50eb-a747-12fbb23038db}{Constellation: Chimera D-Series Gravitational Wave Emission Sourced from Neutrino Anisotropy} the gravitational waveforms sourced by the neutrinos at the same 2664 different observer orientations.]
        A selection of these waveforms (from the same observer orientations as in Fig. \ref{fig:GWs-Flow}) are presented in Fig. \ref{fig:GWs-Neutrino}.
        
        From Fig. \ref{fig:GWs-Neutrino} we see that the amplitudes of the gravitational wave strain sourced by the neutrinos reach similar amplitudes to the strain sourced by the fluid.
        However, due to the non-zero and still evolving neutrino luminosities and their anisotropies, these are not the final strain amplitudes. 
\section{Low-Frequency Gravitational Wave Analysis}
\label{sec:LFGWA}
\begin{figure*}
        \centering
        \includegraphics[width=0.9\linewidth]{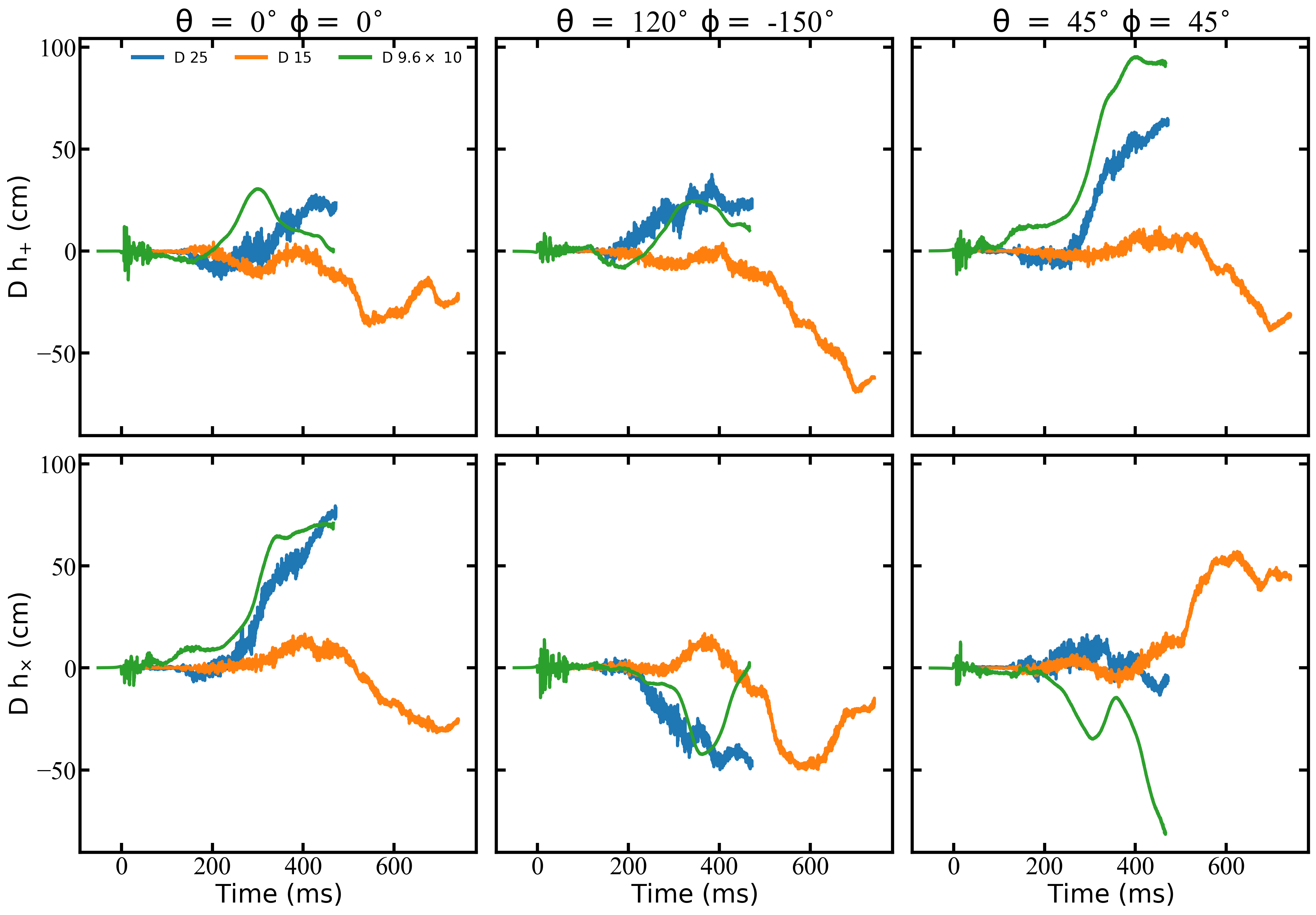}
        \caption{Gravitational waves sourced from \ the fluid and the neutrinos in our core-collapse supernova models. The top row shows the plus polarization at the source. The bottom row shows the cross polarization at the source. Each column shows the different signals at a specific observer orientation with respect to the source frame. Note the significantly lower amplitude signal from the D9.6-3D model, which indicates a more spherical explosion.}
        \label{fig:GWs-Total}
    \end{figure*}

    While the gravitational waves sourced from each species independently can be investigated, in the following analyses we investigate the gravitational waves sourced by all species combined.
    A selection of these waveforms are presented in Fig. \ref{fig:GWs-Total}.
    \subsection{Parameterized Memory Matched Filtering}
    \label{sec:LFGWA-Logistic}
\begin{figure}
            \centering
            \includegraphics[width=0.9\linewidth]{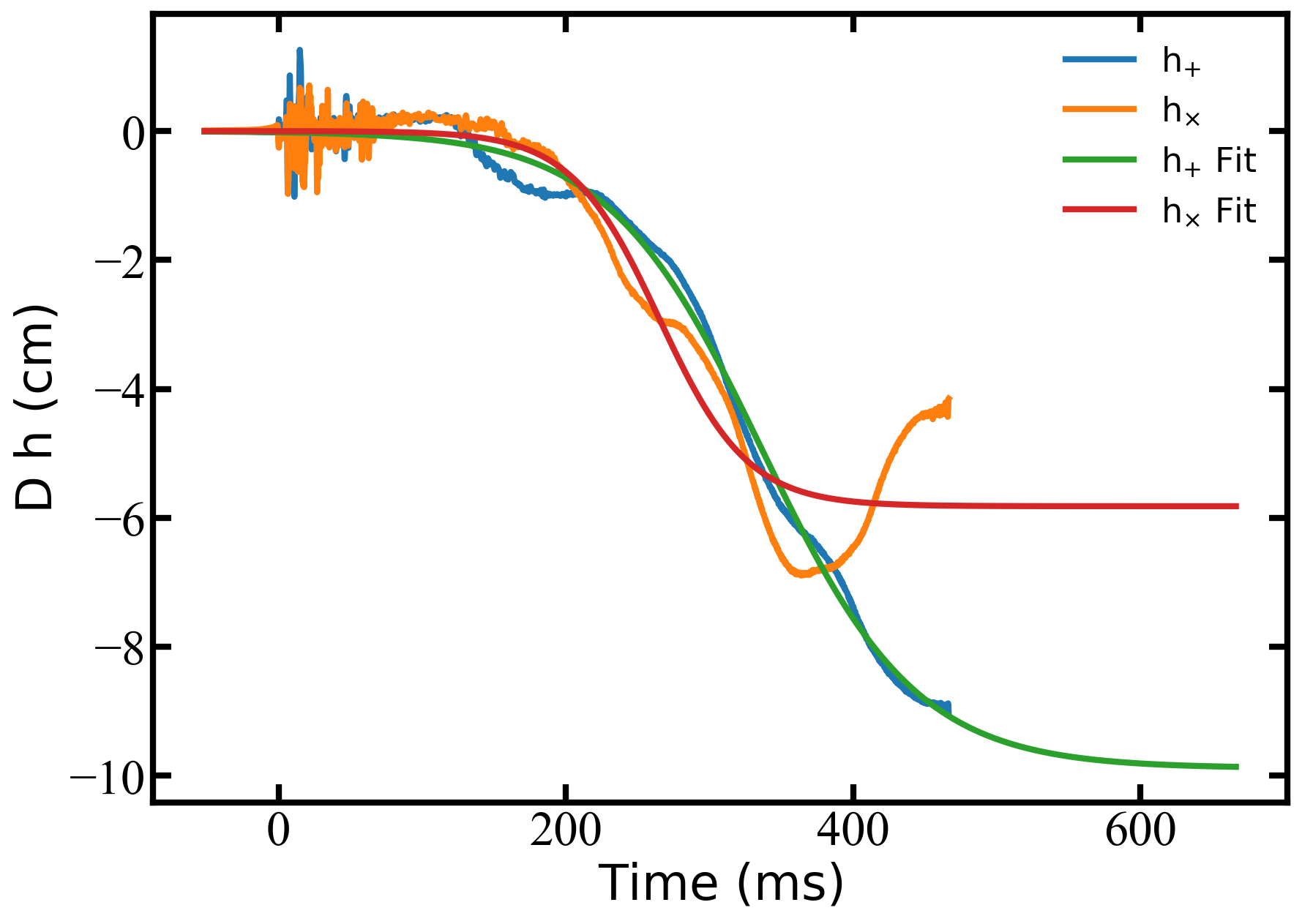}
            \caption{Logistic fit to the total gravitational wave signal from D9.6-3D. For the plus polarization, the values of the fit parameters are $k = 26.70 \pm 0.72$ Hz and $L =  -0.86 \pm 0.008$ cm. For the cross polarization, the fit parameters are $k = 31.23 \pm 0.52$ Hz and $L =  -1.46 \pm  0.005$ cm.}
            \label{fig:D9.6-Total-Fit}
        \end{figure}

\begin{figure}
            \centering
            \includegraphics[width=0.9\linewidth]{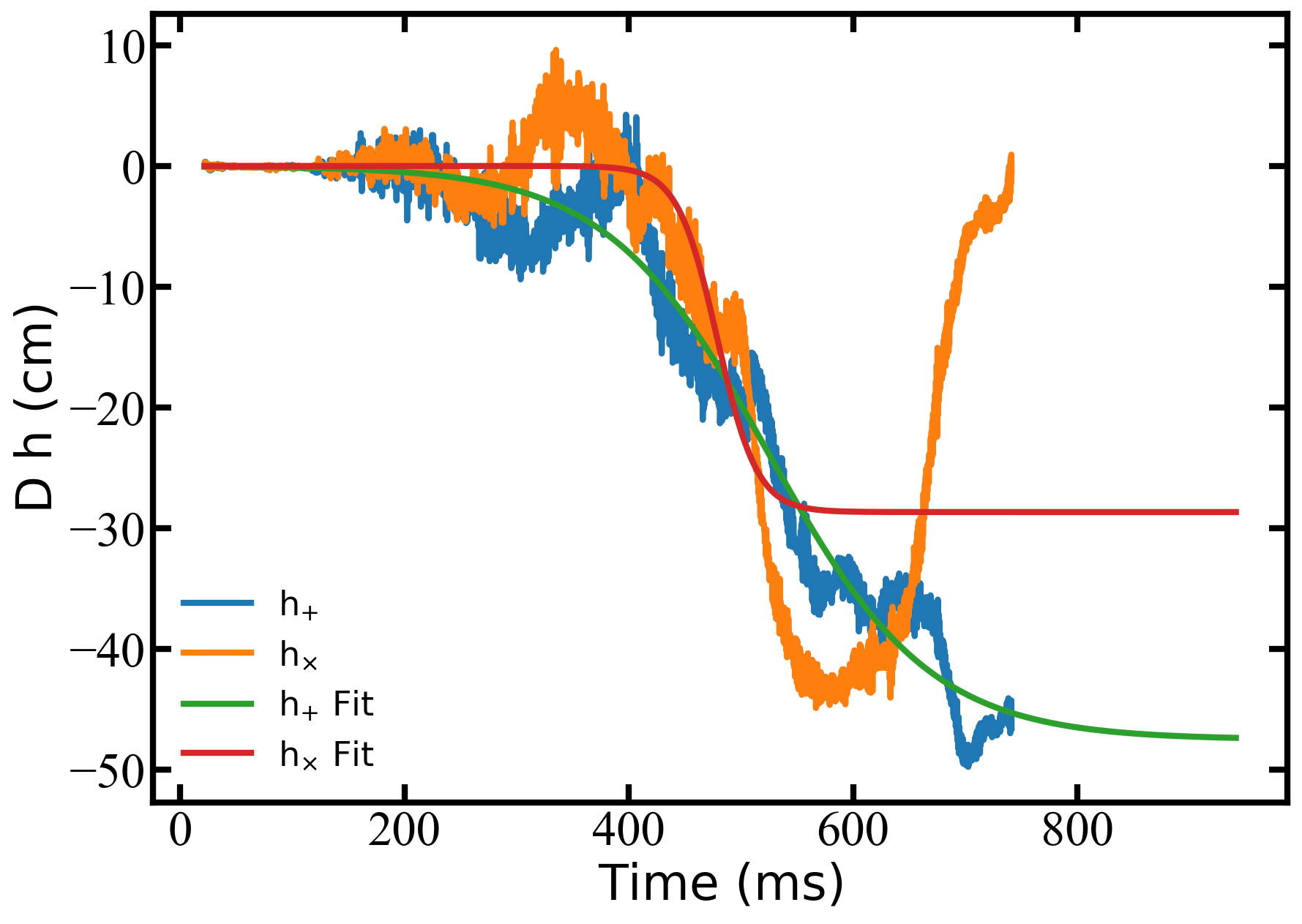}
            \caption{Logistic fit to the total gravitational wave signal from D15-3D. For the plus polarization, the values of the fit parameters are $k = 7.79 \pm 0.0.39$ Hz and $L =   -14.23 \pm 2.34$ cm. For the cross polarization, the fit parameters are $k = 50.86 \pm 2.45$ Hz and $L =  -5.56 \pm 0.04$ cm.}
            \label{fig:D15-Total-Fit}
        \end{figure}

\begin{figure}
            \centering
            \includegraphics[width=0.9\linewidth]{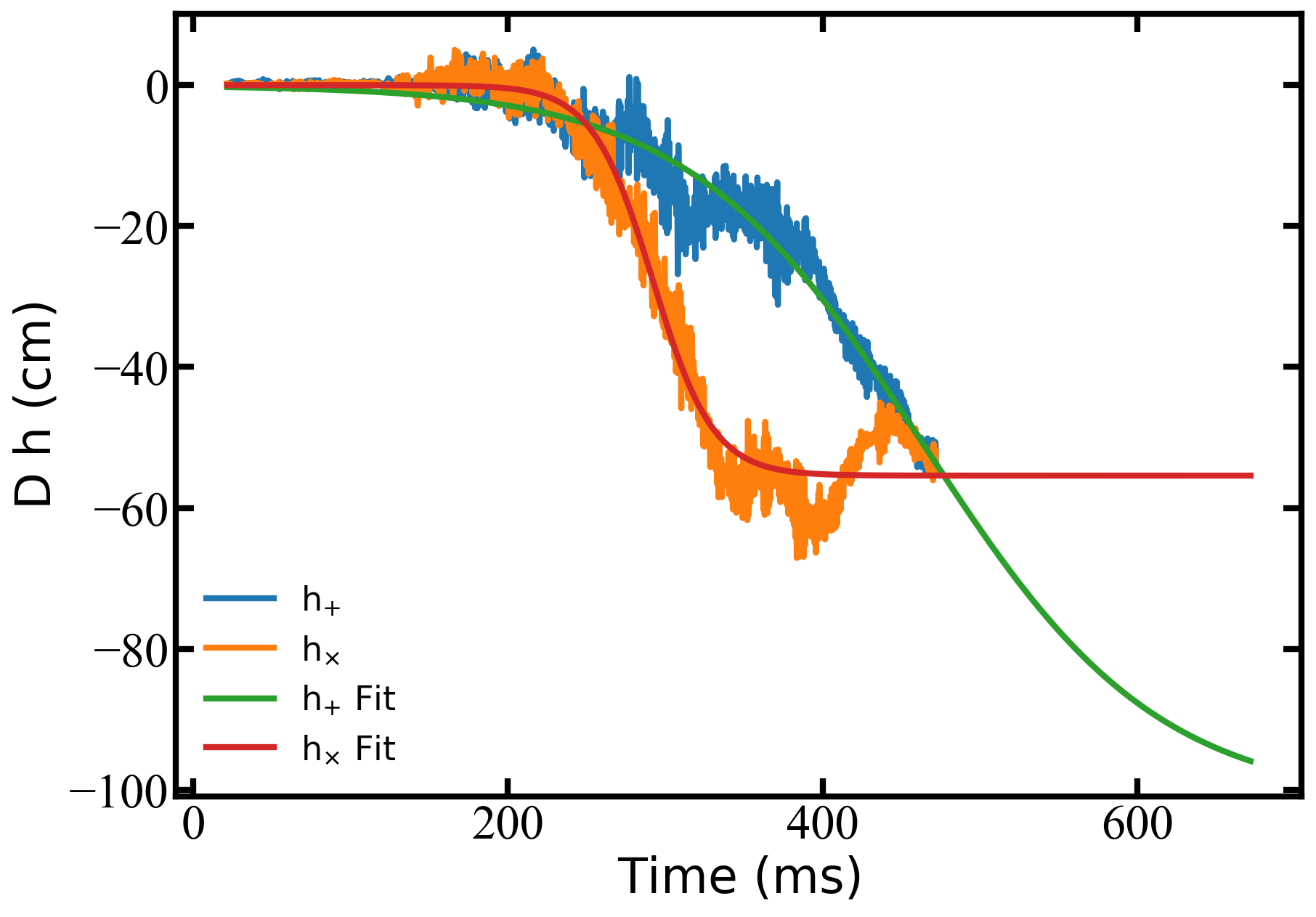}
            \caption{Logistic fit to the total gravitational wave signal from D25-3D. For the plus polarization, the values of the fit parameters are $k = 21.29 \pm 0.66$ Hz and $L =  -18.57 \pm 0.40$ cm. For the cross polarization, the fit parameters are $k = 35.35\pm 3.56$ Hz and $L =  7.57 \pm 0.59$ cm.}
            \label{fig:D25-Total-Fit}
        \end{figure}

        The emission associated with the linear-memory ramp-up, emission below 100 Hz, can be fit with a logistic function, proposed by \citet{2022PhRvD.105j3008R}, given by
            \begin{equation}
            \label{eq:logistic_fit}
                f(t,t_{0},k,L) = \frac{L}{1+e^{-k (t - t_{0})}}.
            \end{equation}
        Here, $t_{0}$ is the center of the rise time, $k$ is the frequency of the memory ramp-up, and $L$ is the memory amplitude. In addition to $k$, we define $\tau = \frac{1}{k}$, which describes the time scale associated with the rise to the memory.
        In Figs. \ref{fig:D9.6-Total-Fit}, \ref{fig:D15-Total-Fit}, and \ref{fig:D25-Total-Fit} we show example fits to the low-frequency signal from models D9.6-3D, D15-3D, and D25-3D, respectively, using the logistic function, Eq. (\ref{eq:logistic_fit}), at an observer angle along $\theta = 100^{\circ}$ and $\phi = -180^{\circ}$. Because the signal depends on the observer orientation, the fitting procedure must be repeated for each orientation. By fitting a logistic function to a large set of observer orientations, we obtain the distribution of the parameters of the fitting function.
        The fact that we can fit the low-frequency component with a relatively simple function allows us to construct templates that we can use to search for the signal using methods similar to the matched filtering methods for compact binaries.
        The low-frequency signals from models D15-3D and D25-3D were still developing when
        the simulations were terminated. The signals have, therefore, not saturated. In the following analysis, we will assume that the signals would continue to develop according to their best fit logistic function.
        \begin{figure*}
            \centering
            \includegraphics[width=0.90\textwidth]{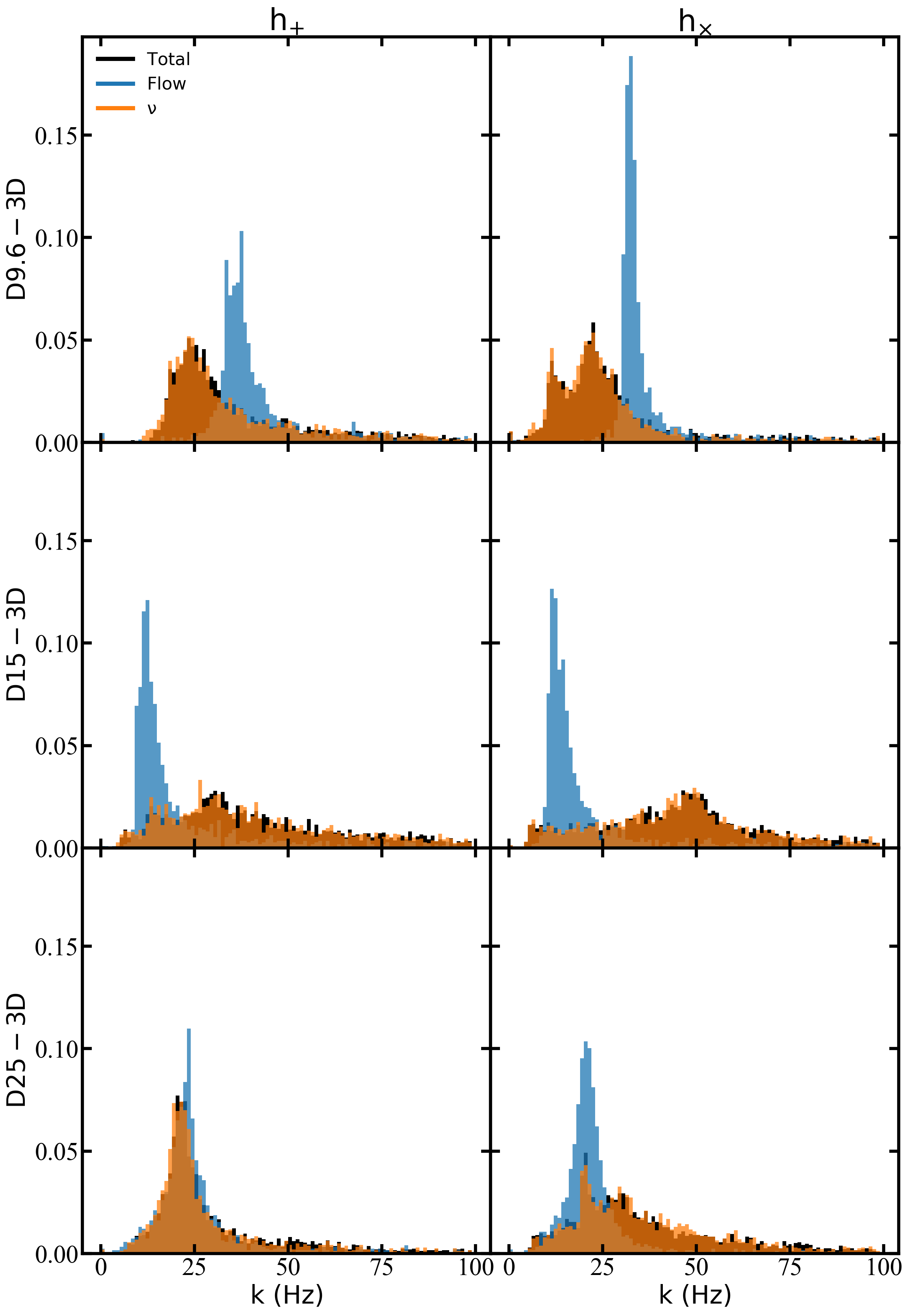}
            \caption{A histogram of the probability density of the fit frequency in the logistic function, as a function of frequency, for each model at different observer orientations.}
            \label{fig:histogram_k}
        \end{figure*}
        \begin{table}[htb]
            \caption{\label{tab:histogram_k_values}Average and standard deviation of the gravitational wave ramp-up frequency (in Hz) from different sources in each model, corresponding to the distributions shown in Fig.~\ref{fig:histogram_k}.}
            \begin{ruledtabular}
                \begin{tabular}{l l c c}
                    Model   & Source     & $\mathrm{D \, h_{+}}$ & $\mathrm{D \, h_{\times}}$ \\
                    \hline
                    D9.6-3D & Flow       & $45.72 \pm 26.69$ & $40.20 \pm 24.16$ \\
                            & Neutrino   & $43.96 \pm 35.73$ & $48.98 \pm 55.976$ \\
                            & Total      & $43.65 \pm 33.98$ & $50.38 \pm 55.68$ \\
                    \hline
                    D15-3D  & Flow       & $41.70 \pm 55.91$ & $49.43 \pm 65.06$ \\
                            & Neutrino   & $60.87 \pm 51.03$ & $56.63 \pm 40.03$ \\
                            & Total      & $61.87 \pm 50.75$ & $55.96 \pm 38.63$ \\
                    \hline
                    D25-3D  & Flow       & $34.63 \pm 39.07$ & $33.88 \pm 43.76$ \\
                            & Neutrino   & $40.98 \pm 43.61$ & $50.75 \pm 45.25$ \\
                            & Total      & $39.58 \pm 38.84$ & $49.61 \pm 44.63$ \\
                \end{tabular}
            \end{ruledtabular}
        \end{table}
        \begin{table}[htb]
            \caption{\label{tab:histogram_L_values}Average and standard deviation of the gravitational wave amplitude (in cm) from different sources in each model.
            }
            \begin{ruledtabular}
                \begin{tabular}{l l c c}
                    Model   & Source     & $\mathrm{D \, h_{+}}$ & $\mathrm{D \, h_{\times}}$ \\
                    \hline
                    D9.6-3D & Flow       & $0.75 \pm 0.41$ & $0.80 \pm 0.46$ \\
                            & Neutrino   & $7.41 \pm 4.06$ & $6.63 \pm 3.78$ \\
                            & Total      & $7.69 \pm 4.23$ & $6.50 \pm 3.73$ \\
                    \hline
                    D15-3D  & Flow       & $6.10 \pm 4.83$   & $5.13 \pm 4.39$ \\
                            & Neutrino   & $31.19 \pm 21.34$ & $32.31 \pm 23.93$ \\
                            & Total      & $30.01 \pm 19.81$ & $32.30 \pm 23.36$ \\
                    \hline
                    D25-3D  & Flow       & $11.04 \pm 6.50$ & $9.75  \pm 6.35$ \\
                            & Neutrino   & $67.20 \pm 53.94$ & $54.94  \pm 51.42$ \\
                            & Total      & $70.34 \pm 57.04$ & $57.84 \pm 54.97$ \\
                \end{tabular}
            \end{ruledtabular}
        \end{table}
        In our logistic-function fit, the frequency, $k$, is the most important parameter. The time $t_{0}$ is of little importance because during matched filtering the template is shifted through time. Therefore, we can effectively redefine the parameter $t_{0}' = t_{0} - t_{\mathrm{central}} = 0$, where $t_{\mathrm{central}}$ is the center of the time-window for the convolution or correlation. The amplitude parameter, $L$, while important for identifying detection prospects for different distances, can, and will, be altered by the signal's injected distance. This leaves the ramp-up frequency, $k$, for which we present in Table \ref{tab:histogram_k_values} the average and the standard deviation, for each distribution shown in Fig. \ref{fig:histogram_k}. From both the figure and the table, we note that the fit performs much better for signals that have stopped evolving (i.e., the D9.6-3D model), providing much tighter distributions (see the bottom left panel of Fig. \ref{fig:GWs-Total}). 
        We also note that while the distribution for the gravitational waves produced by the fluid and the neutrinos are distinct, there is a small amount of overlap, and the distribution for the total gravitational wave signal mimics that of the gravitational waves produced by neutrinos. This further indicates that the low-frequency portion of the total gravitational wave signal is dominated by the contribution from the neutrino anisotropy.
        We also provide a table of the average and standard deviation of the final amplitude $L$ in Table \ref{tab:histogram_L_values}.
        
\begin{figure*}
            \centering
            \includegraphics[width=0.9\linewidth]{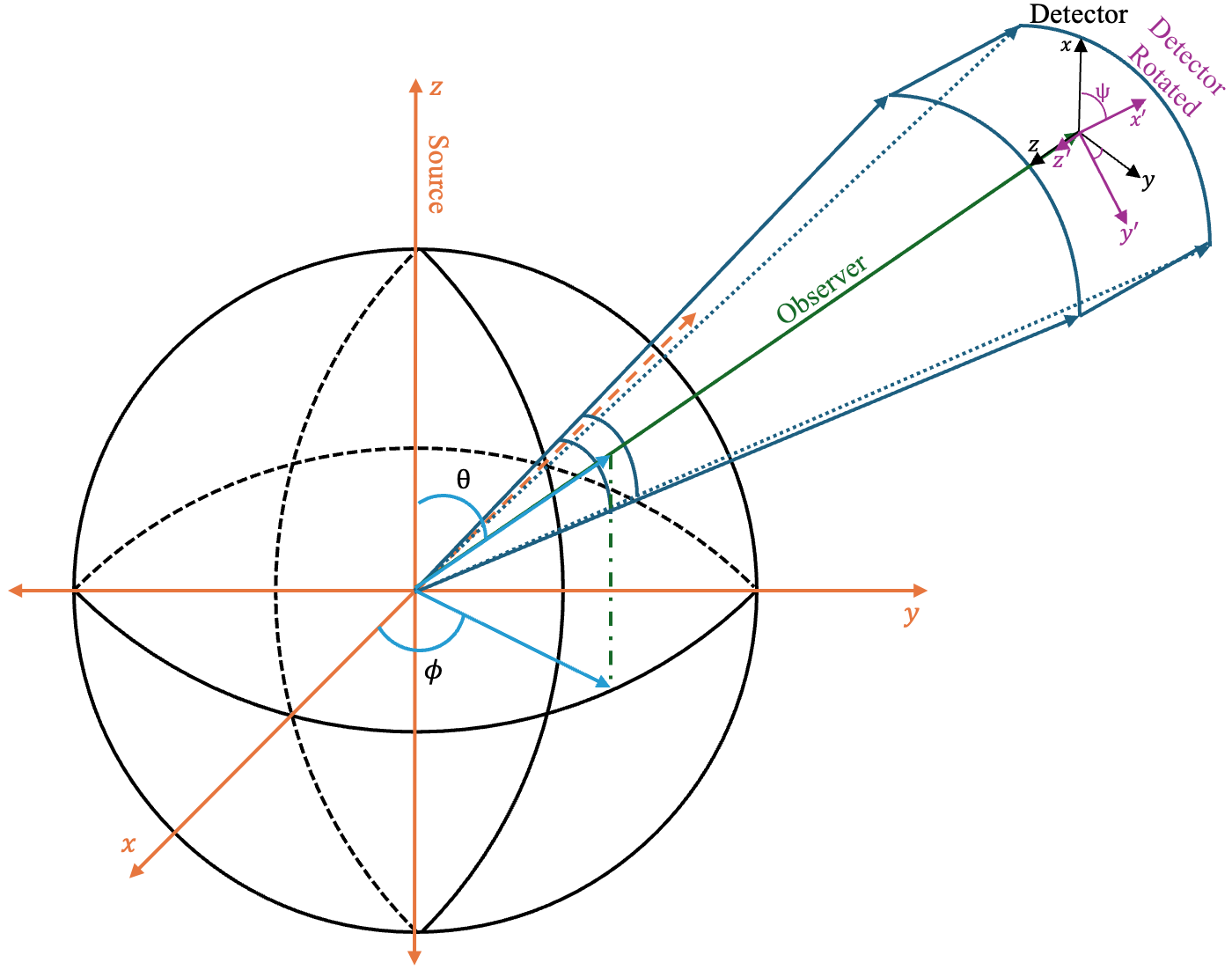}
            \caption{Rotation $\psi$ between the incoming gravitational wave frame and the detector.}
            \label{fig:detector_frame}
        \end{figure*}

        Earlier, we considered all possible detector locations about the source, which resulted in the histograms presented. For a {\em given} detector location, we must also consider all possible orientations of the detector frame (arms) relative to the line of sight to the source. As a first step to include the impact of different relative orientations, we consider rotations about the line of sight in a plane perpendicular to it, as shown in Fig. \ref{fig:detector_frame}. We rotate the strain ``vector,'' whose components correspond to the strain amplitudes for the two polarizations, by an $SO(2)$ transformation of angle $\psi$,
        \begin{equation}
            \begin{bmatrix}
                \tilde{h_{+}} \\
                \tilde{h_{\times}} 
            \end{bmatrix}
             =
             \begin{bmatrix}
                \cos{\psi} & -\sin{\psi} \\
                \sin{\psi} & \cos{\psi} 
            \end{bmatrix}
            \begin{bmatrix}
                h_{+} \\
                h_{\times} 
            \end{bmatrix}.
            \label{eq:rot_pol}
        \end{equation}
         For the signal associated with the D9.6-3D fluid, the only signal no longer evolving, the plus polarization's average frequency is $k = 42.15 \pm 2.03$ Hz and the average amplitude is $L = 0.75 \pm 0.02$ cm. For the cross polarization, we find that the average frequency is $k = 42.15 \pm 2.02$ Hz and the average amplitude is $L = 0.75 \pm 0.02$. The values of the fit parameters for the two polarizations are close to identical. The rotation effectively swaps the polarizations at the maximum rotation angle, $\psi = \pi / 2$, and produces a linear combination of the strains for the two polarizations for all other rotation angles. For completed signals this allows us to define a generic template and a range of parameters to optimize matched-filter searches. 
    \subsection{Reconstruction in Real Interferometric Data}
    \label{sec:LFGWA-cWB}
\begin{figure}
            \centering
            \includegraphics[width=0.9\linewidth]{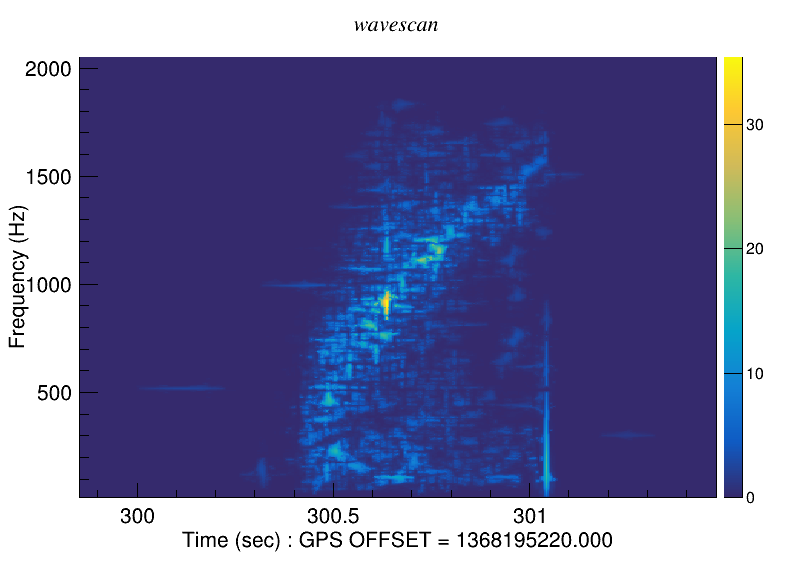}
            \includegraphics[width=0.9\linewidth]{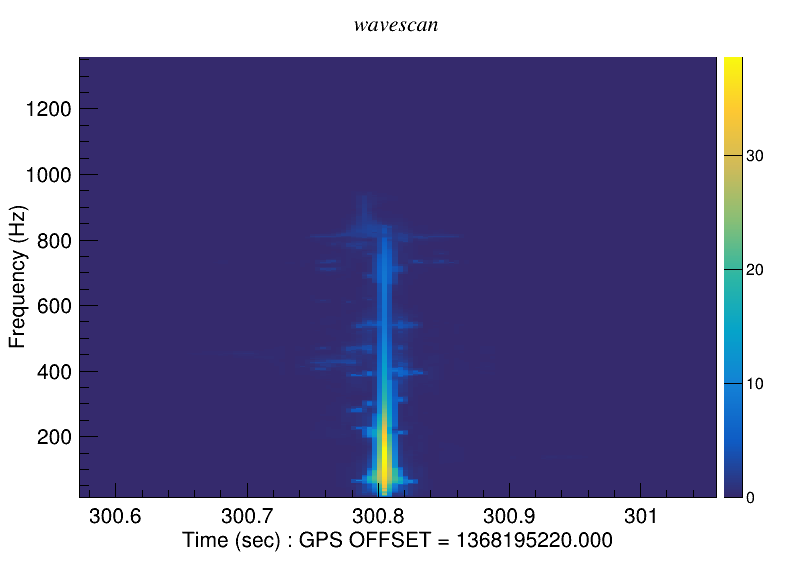}
            \includegraphics[width=0.9\linewidth]{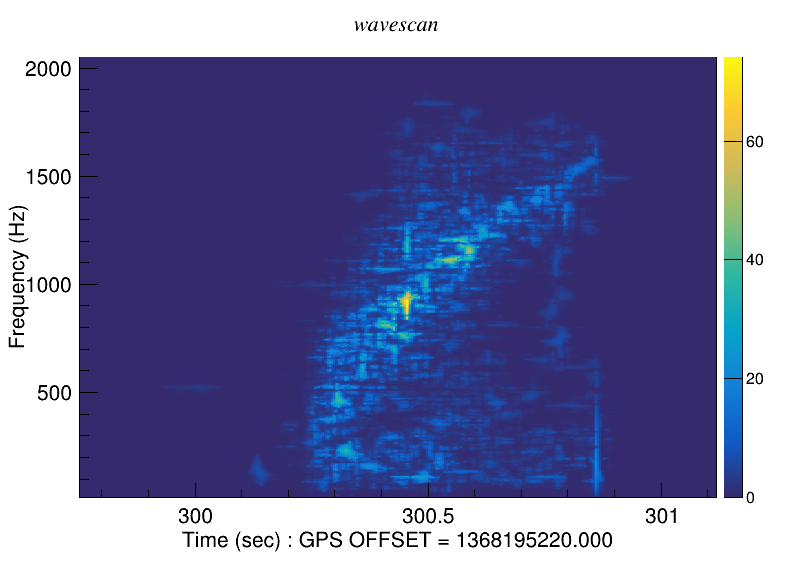}
            \caption{Reconstruction of the gravitational wave signals from D15-3D along the z-axis at 0.1 kpc. The top panel shows the reconstruction of the signal from just the fluid flow. The middle panel shows the reconstruction of the signal from just the neutrino flow. The bottom panel shows the reconstruction of the total signal. }
            \label{fig:cWB_GWs}
        \end{figure}

        In an effort to distinguish the neutrino contribution to the total gravitational wave signal as it arrives at the detectors, utilizing the current detection pipeline, coherent WaveBurst \cite{2016PhRvD..93d2004K, 2021zndo...5798976K}, we inject an example waveform from D15-3D into real interferometric data and attempt to reconstruct the injected signal. There are two motivating factors here: (1) the desire to separate out the neutrino contribution to the gravitational wave signal {\em per se} and (2) that the neutrino contribution dominates the memory. In order to maximize the differences in the fluid, neutrino, and combined signal we inject the signal, at 0.1 kpc (approximately the distance to Betelgeuse). In the fluid case, we see reconstruction of the signal, similar to what is seen in \citet{2024PhRvD.110d2007S}; namely, the reconstruction of the $g/f$-mode feature extending from roughly 400 Hz to 1500 Hz (see the top panel of Fig. \ref{fig:cWB_GWs}). In the case of the gravitational waves generated by the neutrino anisotropy, we do not see any notable portion of the signal (see the middle panel of Fig. \ref{fig:cWB_GWs}). Note that, in this reconstruction, the large ``blip'' near 300.8 seconds is not a feature of the gravitational wave signal, but a high-frequency ringing of the non-tapered injected signal. This ``blip'' disappears when the signal is tapered. However, the results shown here represent the unaltered signal from the CCSN model---i.e., with no tapering, no high-pass filtering, and no other data quality procedures. Finally, the lower panel of Fig. \ref{fig:cWB_GWs} shows the reconstruction of the entire signal. Here we see no contribution from the neutrino-sourced gravitational waves. This shows that for future core-collapse supernova gravitational wave searches using cWB, the portion of the signal from neutrinos does not need to be included.
    \subsection{Detection Prospects}
    \label{sec:LFGWA-Prospects}
        The current detection prospects of gravitational waves from core-collapse supernovae come from aLIGO, aVirgo, and KAGRA \cite{2023ApJS..267...29A}.
        However, a host of proposed detectors are in different stages of planning/building. For future detectors, we limit our study to LISA \cite{2019BAAS...51g..77T}, Cosmic Explorer \cite{2019BAAS...51g..35R}, and the Einstein Telescope \cite{2020JCAP...03..050M}. We highlight LISA, because its milli-Hertz sensitivity range makes it promising for memory searches.
        For both Cosmic Explorer and Einstein Telescope the overall decrease in the noise amplitude and the shifting of the ``low-frequency wall'' to lower frequencies make them promising.  

\begin{figure}
            \includegraphics[width=1.0\linewidth]{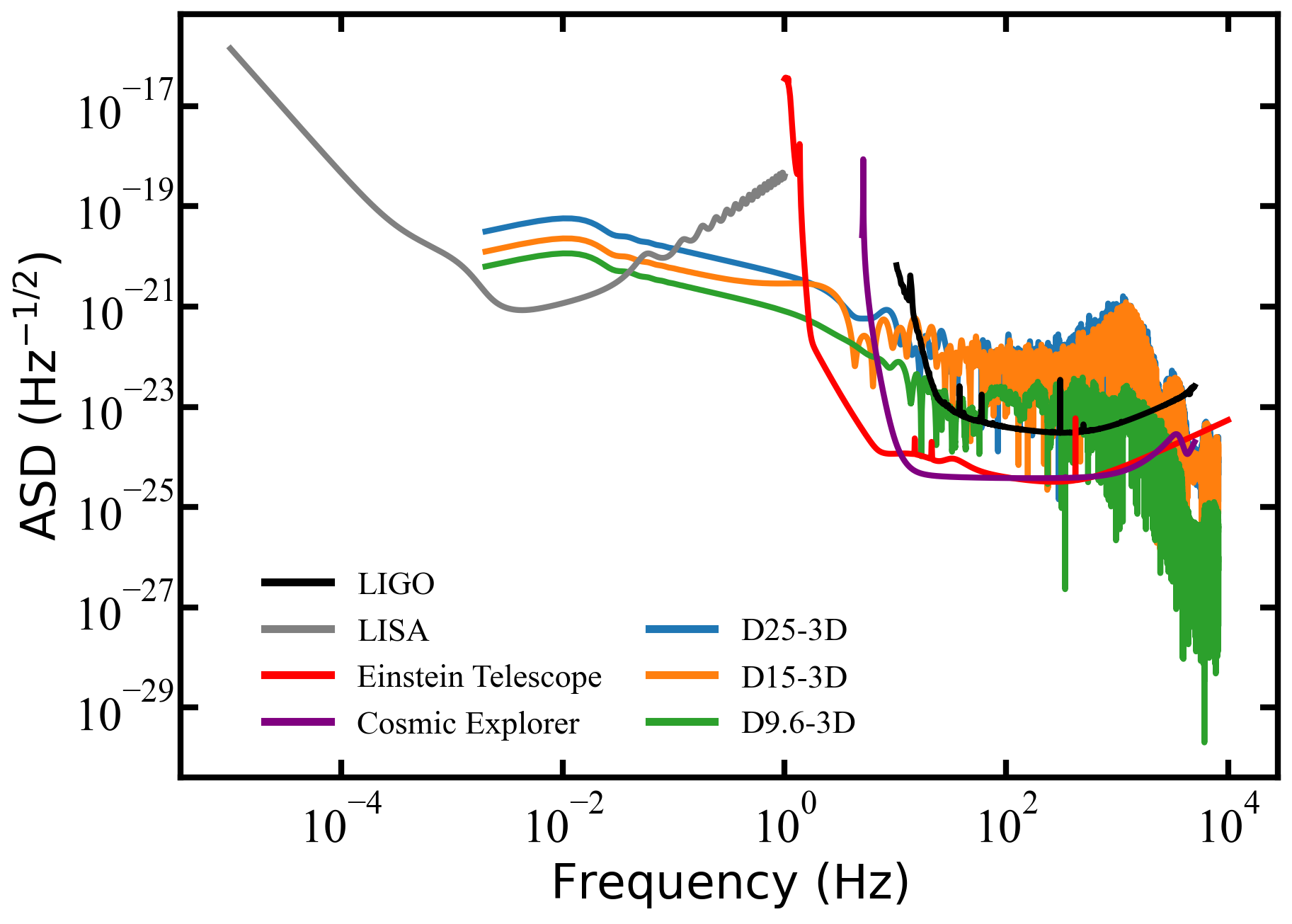}
            \caption{The Amplitude Spectral Density (ASD) of waveforms from our three models at 1 kpc, after a 10000 second extension and tapering, compared to the noise of LIGO's O3 run and the predicted noise of LISA, Cosmic Explorer, and Einstein Telescope.}
            \label{fig:ASD}
        \end{figure}

        The major detectable range for current detectors spans frequencies from approximately 100 Hz to approximately 1000 Hz. However, the noise floors do extend to as low as 10 Hz, before a large increase in noise is encountered. Depending on the time scale of the memory ramp-up and the duration of the memory, we expect some energy to be deposited in the detector, and therefore, a detection to be possible. As shown in \citet{2024PhRvL.133w1401R} for the two LIGO detectors, the use of matched filtering, combined with linear predictive filtering of the noise, predicts memory detection for a Galactic event for sample waveforms from D15-3D and D25-3D. 
        In LISA, we find detectability if the memory signal lasts longer than a few seconds, as seen in the long-duration, flat portion of the signals in the left of Fig. \ref{fig:ASD}, extending above the LISA noise curve. In order to mimic memory time scales, and to remove the high-frequency artifacts from discontinuities in the signal, we add a half-cosine tail that begins at the last strain value and returns to zero smoothly \cite{2022PhRvD.105j3008R},
        \begin{equation}
            h^{\mathrm{extension}}(t) = \frac{L}{2} [1 + \cos{(2 \pi f_{t} (t - t_{s})})],
        \end{equation}
        with $f_{t}$ representing the frequency of the tail and $t_{s}$ representing the time at the end of the simulation.

        Fig. \ref{fig:ASD} shows the amplitude spectral density ($\mathrm{ASD(f)} = |\mathcal{F}(h(t))| * f^{1/2}$) of all three models at 1 kpc, with a 10000 second tail added  (see Fig. \ref{fig:ASD} between $10^{-4}$--$10^{1}$ Hz) specifically to reach the edge of the LISA noise curve. We also note that the change to the signal-to-noise ratio (SNR) is negligible as long as the tapering is longer than the inverse of the left-most intersection point in frequency of the ASD of LISA and the ASD of the signal. At lower frequencies (longer tapering times), the noise is orders of magnitude larger than the signal, and the contribution to the signal-to-noise ratio is negligible.
        As shown in Fig. \ref{fig:ASD}, the intersection point for D9.6-3D is at $\approx 10^{-3}$ Hz, which would indicate a tapering longer than 1000 seconds to maximize the detectability at 1 kpc. The intersection points for D15-3D and D25-3D are at $\approx 5 \times 10^{-4}$ Hz and  $\approx 10^{-4}$ Hz, respectively, which would indicate a tapering longer than 2000 seconds and 10000 seconds, respectively. We extend all three signals to 10000 seconds and calculate the ASDs and SNRs. Given that the gravitational wave amplitude decreases as $\frac{1}{D}$, where D represents the distance in kpc to the source, we can predict detectability out to approximately the center of the Galaxy in the cases of the D15-3D and D25-3D models, assuming a similar detection with SNR $\geq$ 8, as seen in Table \ref{tab:SNR}.
        \begin{table}[htb]
            \caption{\label{tab:SNR}Projected signal-to-noise ratios for our waveforms in predicted noise for LIGO, LISA, Einstein Telescope, and Cosmic Explorer, injected at 1 kiloparsec.}
            \begin{ruledtabular}
                \begin{tabular}{l c c c c}
                Model   & LIGO  & LISA  & ET    & CE    \\ 
                \hline
                D9.6-3D & 13    & 13     & 136   & 91    \\ 
                \hline
                D15-3D  & 94    & 27    & 943   & 1063   \\ 
                \hline
                D25-3D  & 142   & 69    & 1589  & 1269  \\ 
                \end{tabular}
            \end{ruledtabular}
        \end{table}
\section{Conclusions and Discussions}
\label{sec:Conclusion}
    In this paper, we have for the first time presented the gravitational waves sourced from the anisotropy of the neutrino luminosity in the \textsc{Chimera} D9.6-3D, D15-3D, and D25-3D models. We have shown the dependence of the total gravitational wave signal on both the fluid flow and the neutrino flow. In the case of these neutrino-driven events, we highlight that the low-frequency portion of the gravitational wave signal is dominated by the neutrino flow. 
    
    Specifically, in the case of the D9.6-3D model we show that when the explosion has finished and accretion onto the PNS has diminished, the gravitational waves reach a non-zero amplitude, the memory, that persists until the end of the simulation. This allows us to fit a template to the low-frequency signal leading up to and including the memory, in the form of a logistic function with two important parameters, the ``ramp-up'' frequency, $k$, and the final amplitude, $L$. $k$ is of specific importance, reflecting that a component of the evolution to the gravitational wave memory is in a frequency range that is detectable in current detectors, as shown in \cite{2024PhRvL.133w1401R}. By extracting the gravitational wave signals at multiple observer orientations, we are able to determine a characteristic ramp-up frequency and amplitude for each model and each polarization (Tables \ref{tab:histogram_k_values} and \ref{tab:histogram_L_values}, respectively). For the signal from D9.6-3D's fluid flow, the only ``complete'' signal, by introducing a polarization rotation to better represent a gravitational wave signal seen at a detector we show that the characteristic ramp-up frequency is $k_{\mathrm{D9.6-3D}}^{\mathrm{Fluid}} = 42.15 \pm 20.3$ Hz. This result and the preliminary templating shown by the D15-3D and D25-3D signals further motivate a templated search in real interferometric data around 20--60 Hz. 
    The analysis presented here provides us with a template bank that can be used in matched filtering to study detection and parameter estimation. We leave this to future work.

    We utilize coherent WaveBurst  \cite{PhysRevD.93.042004, 2021zndo...5798976K} to reconstruct gravitational wave signals sourced from the fluid flow, neutrino flow, and both, in real interferometric data at 0.1 kpc. We are unable to distinguish any contribution from the neutrinos. 
    As a further study, we seek to utilize a cWB detection in conjunction with a matched-filtering detection to attempt to separate the contributions, by identifying a signal using cWB and then performing matched filtering around the event. In this series analysis, we expect to detect the gravitational waves from the fluid with cWB and the low-frequency portion of the fluid gravitational wave signal, as well as the neutrino gravitational wave signal, using matched filtering. Given the series analysis, we may be able to separate the signals and make predictions about each individual source. This may further assist multi-messenger detections by, for example, utilizing a neutrino detection to search for low-frequency gravitational waves. Following investigations into noise reduction techniques in the 10--30 Hz band, as shown in \citet{2025Sci...389.1012L}, the detection of low-frequency GWs in current interferometers has a promising future. Beyond current detectors, we note the promising future of core-collapse supernova gravitational wave  detection in Cosmic Explorer and Einstein Telescope, where the low-frequency portion of the signal may assist in excess energy detection strategies. We also highlight the future of low-frequency core-collapse supernova gravitational wave detection with the Laser Interferometer Space Antenna, which, given current noise predictions, shows promise. 
    
    The detection of low-frequency gravitational waves from core-collapse supernovae is a promising component of future multimessenger astronomy for which the community must prepare. Studies like this and those indicated here \cite{1978ApJ...223.1037E, 1978Natur.274..565T, 2004ApJ...603..221M, 2009ApJ...697L.133K, 2009ApJ...707.1173M, 2010CQGra..27s4005Y, 2011ApJ...743...30T, 2011ApJ...736..124K, 2012A&A...537A..63M, 2013ApJ...766...43M, 2015PhRvD..92h4040Y, 2018ApJ...861...10M, 2018ApJ...861...10M, 2019MNRAS.487.1178P, 2019ApJ...876L...9R, 2020MNRAS.494.4665P, 2022PhRvD.105j3008R, 2022MNRAS.510.5535J, 2022MNRAS.514.3941N, 2022PhRvD.106d3020M, 2023PhRvD.107d3008M, 2023MNRAS.522.6070P, 2023ApJ...959...21P, 2023PhRvD.107j3015V, 2024PhRvL.133w1401R, 2024ApJ...975...12C, 2021arXiv210505862M, 2024arXiv240513211G} demonstrate progress in the field, progress that must continue.

\begin{acknowledgments}
    C.J.R. would like to acknowledge Jesse Buffaloe and Michael Benjamin for insightful discussions on gravitational wave sourcing and template analysis.
    A.M. acknowledges support from the National Science Foundation's Gravitational Physics Theory Program through grant PHY-2409148.
    H.A. is supported by the Swedish Research Council (Project No. 2020-00452).
    M.Z. is supported by the National Science Foundation Gravitational Physics Experimental and Data Analysis Program through awards PHY-2110555 and PHY-2405227.
    P.M. is supported by the National Science Foundation through its employee IR/D program. The opinions and conclusions expressed herein are those of the authors and do not represent the National Science Foundation.

    An award of computer time was provided by the Innovative and Novel Computational Impact on Theory and Experiment (INCITE) program. This research used resources of the Oak Ridge Leadership Computing Facility, which is a DOE Office of Science User Facility supported under contract DE-AC05-00OR22725.

    The authors would like to acknowledge \citet{2026PhRvD.113h3034L} for pointing out the discrepancy between our computed neutrino-induced waveforms and theirs, which prompted us to investigate the discrepancy and which led to the discovery of our error and corrections herein.

\end{acknowledgments}

\bibliography{Bibliography}%

@PREAMBLE{
 "\providecommand{\noopsort}[1]{}" 
 # "\providecommand{\singleletter}[1]{#1}%" 
}

@ARTICLE{2024PhRvL.133w1401R,
       author = {{Richardson}, Colter J. and {Andresen}, Haakon and {Mezzacappa}, Anthony and {Zanolin}, Michele and {Benjamin}, Michael G. and {Marronetti}, Pedro and {Lentz}, Eric J. and {Szczepa{\'n}czyk}, Marek J.},
        title = "{Detecting Gravitational Wave Memory in the Next Galactic Core-Collapse Supernova}",
      journal = {\prl},
         year = 2024,
        month = dec,
       volume = {133},
       number = {23},
          eid = {231401},
        pages = {231401},
          doi = {10.1103/PhysRevLett.133.231401},
archivePrefix = {arXiv},
       eprint = {2404.02131},
 primaryClass = {astro-ph.HE},
       adsurl = {https://ui.adsabs.harvard.edu/abs/2024PhRvL.133w1401R}
}

@ARTICLE{2022PhRvD.105j3008R,
       author = {{Richardson}, Colter J. and {Zanolin}, Michele and {Andresen}, Haakon and {Szczepa{\'n}czyk}, Marek J. and {Gill}, Kiranjyot and {Wongwathanarat}, Annop},
        title = "{Modeling core-collapse supernovae gravitational-wave memory in laser interferometric data}",
      journal = {\prd},
         year = 2022,
        month = may,
       volume = {105},
       number = {10},
          eid = {103008},
        pages = {103008},
          doi = {10.1103/PhysRevD.105.103008},
archivePrefix = {arXiv},
       eprint = {2109.01582},
 primaryClass = {astro-ph.HE},
       adsurl = {https://ui.adsabs.harvard.edu/abs/2022PhRvD.105j3008R}
}

@ARTICLE{2023PhRvD.107d3008M,
       author = {{Mezzacappa}, Anthony and {Marronetti}, Pedro and {Landfield}, Ryan E. and {Lentz}, Eric J. and {Murphy}, R. Daniel and {Hix}, W. Raphael and {Harris}, J. Austin and {Bruenn}, Stephen W. and {Blondin}, John M. and {Bronson Messer}, O.~E. and {Casanova}, Jordi and {Kronzer}, Luke L.},
        title = "{Core collapse supernova gravitational wave emission for progenitors of 9.6, 15, and 25 M$_{{\ensuremath{\odot}}}$}",
      journal = {\prd},
         year = 2023,
        month = feb,
       volume = {107},
       number = {4},
          eid = {043008},
        pages = {043008},
          doi = {10.1103/PhysRevD.107.043008},
archivePrefix = {arXiv},
       eprint = {2208.10643},
 primaryClass = {astro-ph.SR},
       adsurl = {https://ui.adsabs.harvard.edu/abs/2023PhRvD.107d3008M}
}

@ARTICLE{2010ApJ...724..341H,
       author = {{Heger}, Alexander and {Woosley}, S.~E.},
        title = "{Nucleosynthesis and Evolution of Massive Metal-free Stars}",
      journal = {\apj},
         year = 2010,
        month = nov,
       volume = {724},
       number = {1},
        pages = {341-373},
          doi = {10.1088/0004-637X/724/1/341},
archivePrefix = {arXiv},
       eprint = {0803.3161},
 primaryClass = {astro-ph},
       adsurl = {https://ui.adsabs.harvard.edu/abs/2010ApJ...724..341H}
}

@ARTICLE{2007PhR...442..269W,
       author = {{Woosley}, S.~E. and {Heger}, A.},
        title = "{Nucleosynthesis and remnants in massive stars of solar metallicity}",
      journal = {\physrep},
         year = 2007,
        month = apr,
       volume = {442},
       number = {1-6},
        pages = {269-283},
          doi = {10.1016/j.physrep.2007.02.009},
archivePrefix = {arXiv},
       eprint = {astro-ph/0702176},
 primaryClass = {astro-ph},
       adsurl = {https://ui.adsabs.harvard.edu/abs/2007PhR...442..269W}
}

@article{PhysRevD.93.042004,
  title = {Method for detection and reconstruction of gravitational wave transients with networks of advanced detectors},
  author = {Klimenko, S. and Vedovato, G. and Drago, M. and Salemi, F. and Tiwari, V. and Prodi, G. A. and Lazzaro, C. and Ackley, K. and Tiwari, S. and Da Silva, C. F. and Mitselmakher, G.},
  journal = {Phys. Rev. D},
  volume = {93},
  issue = {4},
  pages = {042004},
  numpages = {10},
  year = {2016},
  month = {Feb},
  publisher = {American Physical Society},
  doi = {10.1103/PhysRevD.93.042004},
  url = {https://link.aps.org/doi/10.1103/PhysRevD.93.042004}
}

@ARTICLE{2012A&A...537A..63M,
       author = {{M{\"u}ller}, E. and {Janka}, H. -Th. and {Wongwathanarat}, A.},
        title = "{Parametrized 3D models of neutrino-driven supernova explosions. Neutrino emission asymmetries and gravitational-wave signals}",
      journal = {\aap},
         year = 2012,
        month = jan,
       volume = {537},
          eid = {A63},
        pages = {A63},
          doi = {10.1051/0004-6361/201117611},
archivePrefix = {arXiv},
       eprint = {1106.6301},
 primaryClass = {astro-ph.SR},
       adsurl = {https://ui.adsabs.harvard.edu/abs/2012A&A...537A..63M}
}

@ARTICLE{2015PhRvD..92h4040Y,
       author = {{Yakunin}, Konstantin N. and {Mezzacappa}, Anthony and {Marronetti}, Pedro and {Yoshida}, Shin'ichirou and {Bruenn}, Stephen W. and {Hix}, W. Raphael and {Lentz}, Eric J. and {Bronson Messer}, O.~E. and {Harris}, J. Austin and {Endeve}, Eirik and {Blondin}, John M. and {Lingerfelt}, Eric J.},
        title = "{Gravitational wave signatures of ab initio two-dimensional core collapse supernova explosion models for 12 -25 M$_{\odot}$ stars}",
      journal = {\prd},
         year = 2015,
        month = oct,
       volume = {92},
       number = {8},
          eid = {084040},
        pages = {084040},
          doi = {10.1103/PhysRevD.92.084040},
archivePrefix = {arXiv},
       eprint = {1505.05824},
 primaryClass = {astro-ph.HE},
       adsurl = {https://ui.adsabs.harvard.edu/abs/2015PhRvD..92h4040Y}
}

@ARTICLE{2016ApJ...829L..14K,
       author = {{Kuroda}, Takami and {Kotake}, Kei and {Takiwaki}, Tomoya},
        title = "{A New Gravitational-wave Signature from Standing Accretion Shock Instability in Supernovae}",
      journal = {\apjl},
         year = 2016,
        month = sep,
       volume = {829},
       number = {1},
          eid = {L14},
        pages = {L14},
          doi = {10.3847/2041-8205/829/1/L14},
archivePrefix = {arXiv},
       eprint = {1605.09215},
 primaryClass = {astro-ph.HE},
       adsurl = {https://ui.adsabs.harvard.edu/abs/2016ApJ...829L..14K}
}

@ARTICLE{2017MNRAS.468.2032A,
       author = {{Andresen}, H. and {M{\"u}ller}, B. and {M{\"u}ller}, E. and {Janka}, H. -Th.},
        title = "{Gravitational wave signals from 3D neutrino hydrodynamics simulations of core-collapse supernovae}",
      journal = {\mnras},
         year = 2017,
        month = jun,
       volume = {468},
       number = {2},
        pages = {2032-2051},
          doi = {10.1093/mnras/stx618},
archivePrefix = {arXiv},
       eprint = {1607.05199},
 primaryClass = {astro-ph.HE},
       adsurl = {https://ui.adsabs.harvard.edu/abs/2017MNRAS.468.2032A}
}

@ARTICLE{2018ApJ...865...81O,
       author = {{O'Connor}, Evan P. and {Couch}, Sean M.},
        title = "{Exploring Fundamentally Three-dimensional Phenomena in High-fidelity Simulations of Core-collapse Supernovae}",
      journal = {\apj},
         year = 2018,
        month = oct,
       volume = {865},
       number = {2},
          eid = {81},
        pages = {81},
          doi = {10.3847/1538-4357/aadcf7},
archivePrefix = {arXiv},
       eprint = {1807.07579},
 primaryClass = {astro-ph.HE},
       adsurl = {https://ui.adsabs.harvard.edu/abs/2018ApJ...865...81O}
}

@ARTICLE{2019MNRAS.486.2238A,
       author = {{Andresen}, H. and {M{\"u}ller}, E. and {Janka}, H. -Th and {Summa}, A. and {Gill}, K. and {Zanolin}, M.},
        title = "{Gravitational waves from 3D core-collapse supernova models: The impact of moderate progenitor rotation}",
      journal = {\mnras},
         year = 2019,
        month = jun,
       volume = {486},
       number = {2},
        pages = {2238-2253},
          doi = {10.1093/mnras/stz990},
archivePrefix = {arXiv},
       eprint = {1810.07638},
 primaryClass = {astro-ph.HE},
       adsurl = {https://ui.adsabs.harvard.edu/abs/2019MNRAS.486.2238A}
}

@ARTICLE{2019MNRAS.487.1178P,
       author = {{Powell}, Jade and {M{\"u}ller}, Bernhard},
        title = "{Gravitational wave emission from 3D explosion models of core-collapse supernovae with low and normal explosion energies}",
      journal = {\mnras},
         year = 2019,
        month = jul,
       volume = {487},
       number = {1},
        pages = {1178-1190},
          doi = {10.1093/mnras/stz1304},
archivePrefix = {arXiv},
       eprint = {1812.05738},
 primaryClass = {astro-ph.HE},
       adsurl = {https://ui.adsabs.harvard.edu/abs/2019MNRAS.487.1178P}
}

@ARTICLE{2019ApJ...876L...9R,
       author = {{Radice}, David and {Morozova}, Viktoriya and {Burrows}, Adam and {Vartanyan}, David and {Nagakura}, Hiroki},
        title = "{Characterizing the Gravitational Wave Signal from Core-collapse Supernovae}",
      journal = {\apjl},
         year = 2019,
        month = may,
       volume = {876},
       number = {1},
          eid = {L9},
        pages = {L9},
          doi = {10.3847/2041-8213/ab191a},
archivePrefix = {arXiv},
       eprint = {1812.07703},
 primaryClass = {astro-ph.HE},
       adsurl = {https://ui.adsabs.harvard.edu/abs/2019ApJ...876L...9R}
}

@ARTICLE{2020MNRAS.494.4665P,
       author = {{Powell}, Jade and {M{\"u}ller}, Bernhard},
        title = "{Three-dimensional core-collapse supernova simulations of massive and rotating progenitors}",
      journal = {\mnras},
         year = 2020,
        month = jun,
       volume = {494},
       number = {4},
        pages = {4665-4675},
          doi = {10.1093/mnras/staa1048},
archivePrefix = {arXiv},
       eprint = {2002.10115},
 primaryClass = {astro-ph.HE},
       adsurl = {https://ui.adsabs.harvard.edu/abs/2020MNRAS.494.4665P}
}

@ARTICLE{2020PhRvD.102b3027M,
       author = {{Mezzacappa}, Anthony and {Marronetti}, Pedro and {Landfield}, Ryan E. and {Lentz}, Eric J. and {Yakunin}, Konstantin N. and {Bruenn}, Stephen W. and {Hix}, W. Raphael and {Messer}, O.~E. Bronson and {Endeve}, Eirik and {Blondin}, John M. and {Harris}, J. Austin},
        title = "{Gravitational-wave signal of a core-collapse supernova explosion of a 15 M$_{\odot}$ star}",
      journal = {\prd},
         year = 2020,
        month = jul,
       volume = {102},
       number = {2},
          eid = {023027},
        pages = {023027},
          doi = {10.1103/PhysRevD.102.023027},
archivePrefix = {arXiv},
       eprint = {2007.15099},
 primaryClass = {astro-ph.HE},
       adsurl = {https://ui.adsabs.harvard.edu/abs/2020PhRvD.102b3027M}
}

@ARTICLE{2021MNRAS.503.3552A,
       author = {{Andresen}, H. and {Glas}, R. and {Janka}, H. -Th},
        title = "{Gravitational-wave signals from 3D supernova simulations with different neutrino-transport methods}",
      journal = {\mnras},
         year = 2021,
        month = may,
       volume = {503},
       number = {3},
        pages = {3552-3567},
          doi = {10.1093/mnras/stab675},
archivePrefix = {arXiv},
       eprint = {2011.10499},
 primaryClass = {astro-ph.HE},
       adsurl = {https://ui.adsabs.harvard.edu/abs/2021MNRAS.503.3552A}
      }

@ARTICLE{2020ApJS..248...11B,
       author = {{Bruenn}, Stephen W. and {Blondin}, John M. and {Hix}, W. Raphael and {Lentz}, Eric J. and {Messer}, O.~E. Bronson and {Mezzacappa}, Anthony and {Endeve}, Eirik and {Harris}, J. Austin and {Marronetti}, Pedro and {Budiardja}, Reuben D. and {Chertkow}, Merek A. and {Lee}, Ching-Tsai},
        title = "{CHIMERA: A Massively Parallel Code for Core-collapse Supernova Simulations}",
      journal = {\apjs},
         year = 2020,
        month = may,
       volume = {248},
       number = {1},
          eid = {11},
        pages = {11},
          doi = {10.3847/1538-4365/ab7aff},
archivePrefix = {arXiv},
       eprint = {1809.05608},
 primaryClass = {astro-ph.IM},
       adsurl = {https://ui.adsabs.harvard.edu/abs/2020ApJS..248...11B}
}

@ARTICLE{2023ApJS..267...29A,
       author = {{The LIGO Scientific Collaboration} and {the Virgo Collaboration} and {the KAGRA Collaboration}},
        title = "{Open Data from the Third Observing Run of LIGO, Virgo, KAGRA, and GEO}",
      journal = {\apjs},
         year = 2023,
        month = aug,
       volume = {267},
       number = {2},
          eid = {29},
        pages = {29},
          doi = {10.3847/1538-4365/acdc9f},
archivePrefix = {arXiv},
       eprint = {2302.03676},
 primaryClass = {gr-qc},
       adsurl = {https://ui.adsabs.harvard.edu/abs/2023ApJS..267...29A}
}

@ARTICLE{2021ApJ...921..113S,
       author = {{Sandoval}, Michael A. and {Hix}, W. Raphael and {Messer}, O.~E. Bronson and {Lentz}, Eric J. and {Harris}, J. Austin},
        title = "{Three-dimensional Core-collapse Supernova Simulations with 160 Isotopic Species Evolved to Shock Breakout}",
      journal = {\apj},
         year = 2021,
        month = nov,
       volume = {921},
       number = {2},
          eid = {113},
        pages = {113},
          doi = {10.3847/1538-4357/ac1d49},
archivePrefix = {arXiv},
       eprint = {2106.01389},
 primaryClass = {astro-ph.HE},
       adsurl = {https://ui.adsabs.harvard.edu/abs/2021ApJ...921..113S}
}

@ARTICLE{1978ApJ...223.1037E,
       author = {{Epstein}, R.},
        title = "{The generation of gravitational radiation by escaping supernova neutrinos.}",
      journal = {\apj},
         year = 1978,
        month = aug,
       volume = {223},
        pages = {1037-1045},
          doi = {10.1086/156337},
       adsurl = {https://ui.adsabs.harvard.edu/abs/1978ApJ...223.1037E}
}

@BOOK{1973grav.book.....M,
       author = {{Misner}, Charles W. and {Thorne}, Kip S. and {Wheeler}, John Archibald},
        title = "{Gravitation}",
         year = 1973,
       adsurl = {https://ui.adsabs.harvard.edu/abs/1973grav.book.....M}
}

@ARTICLE{2025arXiv250306406M,
       author = {{Murphy}, R. Daniel and {Mezzacappa}, Anthony and {Lentz}, Eric J. and {Marronetti}, Pedro},
        title = "{Core Collapse Supernova Gravitational Wave Sourcing and Characterization based on Three-Dimensional Models}",
      journal = {arXiv e-prints},
         year = 2025,
        month = mar,
          eid = {arXiv:2503.06406},
        pages = {arXiv:2503.06406},
          doi = {10.48550/arXiv.2503.06406},
archivePrefix = {arXiv},
       eprint = {2503.06406},
 primaryClass = {astro-ph.HE},
       adsurl = {https://ui.adsabs.harvard.edu/abs/2025arXiv250306406M}
}

@ARTICLE{2024PhRvD.110h3006M,
       author = {{Murphy}, R. Daniel and {Casallas-Lagos}, Alejandro and {Mezzacappa}, Anthony and {Zanolin}, Michele and {Landfield}, Ryan E. and {Lentz}, Eric J. and {Marronetti}, Pedro and {Antelis}, Javier M. and {Moreno}, Claudia},
        title = "{Dependence of the reconstructed core-collapse supernova gravitational wave high-frequency feature on the nuclear equation of state in real interferometric data}",
      journal = {\prd},
         year = 2024,
        month = oct,
       volume = {110},
       number = {8},
          eid = {083006},
        pages = {083006},
          doi = {10.1103/PhysRevD.110.083006},
archivePrefix = {arXiv},
       eprint = {2406.01784},
 primaryClass = {astro-ph.HE},
       adsurl = {https://ui.adsabs.harvard.edu/abs/2024PhRvD.110h3006M}
}

@ARTICLE{1982A&A...114...53M,
       author = {{Mueller}, E.},
        title = "{Gravitational radiation from collapsing rotating stellar cores}",
      journal = {\aap},
         year = 1982,
        month = oct,
       volume = {114},
       number = {1},
        pages = {53-59},
       adsurl = {https://ui.adsabs.harvard.edu/abs/1982A&A...114...53M}
}

@ARTICLE{1991A&A...246..417M,
       author = {{Moenchmeyer}, R. and {Schaefer}, G. and {Mueller}, E. and {Kates}, R.~E.},
        title = "{Gravitational waves from the collapse of rotating stellar cores.}",
      journal = {\aap},
         year = 1991,
        month = jun,
       volume = {246},
        pages = {417-440},
       adsurl = {https://ui.adsabs.harvard.edu/abs/1991A&A...246..417M}
}

@ARTICLE{1997A&A...317..140M,
       author = {{Mueller}, E. and {Janka}, H. -T.},
        title = "{Gravitational radiation from convective instabilities in Type II supernova explosions.}",
      journal = {\aap},
         year = 1997,
        month = jan,
       volume = {317},
        pages = {140-163},
       adsurl = {https://ui.adsabs.harvard.edu/abs/1997A&A...317..140M}
}

@ARTICLE{2001ApJ...560L.163D,
       author = {{Dimmelmeier}, Harald and {Font}, Jos{\'e} A. and {M{\"u}ller}, Ewald},
        title = "{Gravitational Waves from Relativistic Rotational Core Collapse}",
      journal = {\apjl},
         year = 2001,
        month = oct,
       volume = {560},
       number = {2},
        pages = {L163-L166},
          doi = {10.1086/324406},
archivePrefix = {arXiv},
       eprint = {astro-ph/0103088},
 primaryClass = {astro-ph},
       adsurl = {https://ui.adsabs.harvard.edu/abs/2001ApJ...560L.163D}
}

@ARTICLE{2003PhRvD..68d4023K,
       author = {{Kotake}, Kei and {Yamada}, Shoichi and {Sato}, Katsuhiko},
        title = "{Gravitational radiation from axisymmetric rotational core collapse}",
      journal = {\prd},
         year = 2003,
        month = aug,
       volume = {68},
       number = {4},
          eid = {044023},
        pages = {044023},
          doi = {10.1103/PhysRevD.68.044023},
archivePrefix = {arXiv},
       eprint = {astro-ph/0306430},
 primaryClass = {astro-ph},
       adsurl = {https://ui.adsabs.harvard.edu/abs/2003PhRvD..68d4023K}
}

@ARTICLE{2004PhRvD..69l4004K,
       author = {{Kotake}, Kei and {Yamada}, Shoichi and {Sato}, Katsuhiko and {Sumiyoshi}, Kohsuke and {Ono}, Hiroyuki and {Suzuki}, Hideyuki},
        title = "{Gravitational radiation from rotational core collapse: Effects of magnetic fields and realistic equations of state}",
      journal = {\prd},
         year = 2004,
        month = jun,
       volume = {69},
       number = {12},
          eid = {124004},
        pages = {124004},
          doi = {10.1103/PhysRevD.69.124004},
archivePrefix = {arXiv},
       eprint = {astro-ph/0401563},
 primaryClass = {astro-ph},
       adsurl = {https://ui.adsabs.harvard.edu/abs/2004PhRvD..69l4004K}
}

@ARTICLE{2004ApJ...603..221M,
       author = {{M{\"u}ller}, Ewald and {Rampp}, Markus and {Buras}, Robert and {Janka}, H. -Thomas and {Shoemaker}, David H.},
        title = "{Toward Gravitational Wave Signals from Realistic Core-Collapse Supernova Models}",
      journal = {\apj},
         year = 2004,
        month = mar,
       volume = {603},
       number = {1},
        pages = {221-230},
          doi = {10.1086/381360},
archivePrefix = {arXiv},
       eprint = {astro-ph/0309833},
 primaryClass = {astro-ph},
       adsurl = {https://ui.adsabs.harvard.edu/abs/2004ApJ...603..221M}
}

@ARTICLE{2009ApJ...697L.133K,
       author = {{Kotake}, Kei and {Iwakami}, Wakana and {Ohnishi}, Naofumi and {Yamada}, Shoichi},
        title = "{Stochastic Nature of Gravitational Waves from Supernova Explosions with Standing Accretion Shock Instability}",
      journal = {\apjl},
         year = 2009,
        month = jun,
       volume = {697},
       number = {2},
        pages = {L133-L136},
          doi = {10.1088/0004-637X/697/2/L133},
archivePrefix = {arXiv},
       eprint = {0904.4300},
 primaryClass = {astro-ph.HE},
       adsurl = {https://ui.adsabs.harvard.edu/abs/2009ApJ...697L.133K}
}

@ARTICLE{2009A&A...496..475M,
       author = {{Marek}, A. and {Janka}, H. -T. and {M{\"u}ller}, E.},
        title = "{Equation-of-state dependent features in shock-oscillation modulated neutrino and gravitational-wave signals from supernovae}",
      journal = {\aap},
         year = 2009,
        month = mar,
       volume = {496},
       number = {2},
        pages = {475-494},
          doi = {10.1051/0004-6361/200810883},
archivePrefix = {arXiv},
       eprint = {0808.4136},
 primaryClass = {astro-ph},
       adsurl = {https://ui.adsabs.harvard.edu/abs/2009A&A...496..475M}
}

@ARTICLE{2009ApJ...707.1173M,
       author = {{Murphy}, Jeremiah W. and {Ott}, Christian D. and {Burrows}, Adam},
        title = "{A Model for Gravitational Wave Emission from Neutrino-Driven Core-Collapse Supernovae}",
      journal = {\apj},
         year = 2009,
        month = dec,
       volume = {707},
       number = {2},
        pages = {1173-1190},
          doi = {10.1088/0004-637X/707/2/1173},
archivePrefix = {arXiv},
       eprint = {0907.4762},
 primaryClass = {astro-ph.SR},
       adsurl = {https://ui.adsabs.harvard.edu/abs/2009ApJ...707.1173M}
}

@ARTICLE{2010A&A...514A..51S,
       author = {{Scheidegger}, S. and {K{\"a}ppeli}, R. and {Whitehouse}, S.~C. and {Fischer}, T. and {Liebend{\"o}rfer}, M.},
        title = "{The influence of model parameters on the prediction of gravitational wave signals from stellar core collapse}",
      journal = {\aap},
         year = 2010,
        month = may,
       volume = {514},
          eid = {A51},
        pages = {A51},
          doi = {10.1051/0004-6361/200913220},
archivePrefix = {arXiv},
       eprint = {1001.1570},
 primaryClass = {astro-ph.HE},
       adsurl = {https://ui.adsabs.harvard.edu/abs/2010A&A...514A..51S}
}

@ARTICLE{2011ApJ...736..124K,
       author = {{Kotake}, Kei and {Iwakami-Nakano}, Wakana and {Ohnishi}, Naofumi},
        title = "{Effects of Rotation on Stochasticity of Gravitational Waves in the Nonlinear Phase of Core-collapse Supernovae}",
      journal = {\apj},
         year = 2011,
        month = aug,
       volume = {736},
       number = {2},
          eid = {124},
        pages = {124},
          doi = {10.1088/0004-637X/736/2/124},
archivePrefix = {arXiv},
       eprint = {1106.0544},
 primaryClass = {astro-ph.HE},
       adsurl = {https://ui.adsabs.harvard.edu/abs/2011ApJ...736..124K}
}

@ARTICLE{2013ApJ...779L..18C,
       author = {{Cerd{\'a}-Dur{\'a}n}, Pablo and {DeBrye}, Nicolas and {Aloy}, Miguel A. and {Font}, Jos{\'e} A. and {Obergaulinger}, Martin},
        title = "{Gravitational Wave Signatures in Black Hole Forming Core Collapse}",
      journal = {\apjl},
         year = 2013,
        month = dec,
       volume = {779},
       number = {2},
          eid = {L18},
        pages = {L18},
          doi = {10.1088/2041-8205/779/2/L18},
archivePrefix = {arXiv},
       eprint = {1310.8290},
 primaryClass = {astro-ph.SR},
       adsurl = {https://ui.adsabs.harvard.edu/abs/2013ApJ...779L..18C}
}

@ARTICLE{2013ApJ...766...43M,
       author = {{M{\"u}ller}, Bernhard and {Janka}, Hans-Thomas and {Marek}, Andreas},
        title = "{A New Multi-dimensional General Relativistic Neutrino Hydrodynamics Code of Core-collapse Supernovae. III. Gravitational Wave Signals from Supernova Explosion Models}",
      journal = {\apj},
         year = 2013,
        month = mar,
       volume = {766},
       number = {1},
          eid = {43},
        pages = {43},
          doi = {10.1088/0004-637X/766/1/43},
archivePrefix = {arXiv},
       eprint = {1210.6984},
 primaryClass = {astro-ph.SR},
       adsurl = {https://ui.adsabs.harvard.edu/abs/2013ApJ...766...43M}
}

@ARTICLE{2013ApJ...768..115O,
       author = {{Ott}, Christian D. and {Abdikamalov}, Ernazar and {M{\"o}sta}, Philipp and {Haas}, Roland and {Drasco}, Steve and {O'Connor}, Evan P. and {Reisswig}, Christian and {Meakin}, Casey A. and {Schnetter}, Erik},
        title = "{General-relativistic Simulations of Three-dimensional Core-collapse Supernovae}",
      journal = {\apj},
         year = 2013,
        month = may,
       volume = {768},
       number = {2},
          eid = {115},
        pages = {115},
          doi = {10.1088/0004-637X/768/2/115},
archivePrefix = {arXiv},
       eprint = {1210.6674},
 primaryClass = {astro-ph.HE},
       adsurl = {https://ui.adsabs.harvard.edu/abs/2013ApJ...768..115O}
}

@ARTICLE{2014PhRvD..89d4011K,
       author = {{Kuroda}, Takami and {Takiwaki}, Tomoya and {Kotake}, Kei},
        title = "{Gravitational wave signatures from low-mode spiral instabilities in rapidly rotating supernova cores}",
      journal = {\prd},
         year = 2014,
        month = feb,
       volume = {89},
       number = {4},
          eid = {044011},
        pages = {044011},
          doi = {10.1103/PhysRevD.89.044011},
archivePrefix = {arXiv},
       eprint = {1304.4372},
 primaryClass = {astro-ph.HE},
       adsurl = {https://ui.adsabs.harvard.edu/abs/2014PhRvD..89d4011K}
}

@ARTICLE{2015PhRvD..92l2001H,
       author = {{Hayama}, Kazuhiro and {Kuroda}, Takami and {Kotake}, Kei and {Takiwaki}, Tomoya},
        title = "{Coherent network analysis of gravitational waves from three-dimensional core-collapse supernova models}",
      journal = {\prd},
         year = 2015,
        month = dec,
       volume = {92},
       number = {12},
          eid = {122001},
        pages = {122001},
          doi = {10.1103/PhysRevD.92.122001},
archivePrefix = {arXiv},
       eprint = {1501.00966},
 primaryClass = {astro-ph.HE},
       adsurl = {https://ui.adsabs.harvard.edu/abs/2015PhRvD..92l2001H}
}

@ARTICLE{2016PhRvL.116o1102H,
       author = {{Hayama}, Kazuhiro and {Kuroda}, Takami and {Nakamura}, Ko and {Yamada}, Shoichi},
        title = "{Circular Polarizations of Gravitational Waves from Core-Collapse Supernovae: A Clear Indication of Rapid Rotation}",
      journal = {\prl},
         year = 2016,
        month = apr,
       volume = {116},
       number = {15},
          eid = {151102},
        pages = {151102},
          doi = {10.1103/PhysRevLett.116.151102},
archivePrefix = {arXiv},
       eprint = {1606.01520},
 primaryClass = {astro-ph.HE},
       adsurl = {https://ui.adsabs.harvard.edu/abs/2016PhRvL.116o1102H}
}

@ARTICLE{2017ApJ...851...62K,
       author = {{Kuroda}, Takami and {Kotake}, Kei and {Hayama}, Kazuhiro and {Takiwaki}, Tomoya},
        title = "{Correlated Signatures of Gravitational-wave and Neutrino Emission in Three-dimensional General-relativistic Core-collapse Supernova Simulations}",
      journal = {\apj},
         year = 2017,
        month = dec,
       volume = {851},
       number = {1},
          eid = {62},
        pages = {62},
          doi = {10.3847/1538-4357/aa988d},
archivePrefix = {arXiv},
       eprint = {1708.05252},
 primaryClass = {astro-ph.HE},
       adsurl = {https://ui.adsabs.harvard.edu/abs/2017ApJ...851...62K}
}

@ARTICLE{2017PhRvD..95f3019R,
       author = {{Richers}, Sherwood and {Ott}, Christian D. and {Abdikamalov}, Ernazar and {O'Connor}, Evan and {Sullivan}, Chris},
        title = "{Equation of state effects on gravitational waves from rotating core collapse}",
      journal = {\prd},
         year = 2017,
        month = mar,
       volume = {95},
       number = {6},
          eid = {063019},
        pages = {063019},
          doi = {10.1103/PhysRevD.95.063019},
archivePrefix = {arXiv},
       eprint = {1701.02752},
 primaryClass = {astro-ph.HE},
       adsurl = {https://ui.adsabs.harvard.edu/abs/2017PhRvD..95f3019R}
}

@ARTICLE{2018ApJ...861...10M,
       author = {{Morozova}, Viktoriya and {Radice}, David and {Burrows}, Adam and {Vartanyan}, David},
        title = "{The Gravitational Wave Signal from Core-collapse Supernovae}",
      journal = {\apj},
         year = 2018,
        month = jul,
       volume = {861},
       number = {1},
          eid = {10},
        pages = {10},
          doi = {10.3847/1538-4357/aac5f1},
archivePrefix = {arXiv},
       eprint = {1801.01914},
 primaryClass = {astro-ph.HE},
       adsurl = {https://ui.adsabs.harvard.edu/abs/2018ApJ...861...10M}
}

@ARTICLE{2018MNRAS.475L..91T,
       author = {{Takiwaki}, Tomoya and {Kotake}, Kei},
        title = "{Anisotropic emission of neutrino and gravitational-wave signals from rapidly rotating core-collapse supernovae}",
      journal = {\mnras},
         year = 2018,
        month = mar,
       volume = {475},
       number = {1},
        pages = {L91-L95},
          doi = {10.1093/mnrasl/sly008},
archivePrefix = {arXiv},
       eprint = {1711.01905},
 primaryClass = {astro-ph.HE},
       adsurl = {https://ui.adsabs.harvard.edu/abs/2018MNRAS.475L..91T}
}

@ARTICLE{2018MNRAS.477L..96H,
       author = {{Hayama}, Kazuhiro and {Kuroda}, Takami and {Kotake}, Kei and {Takiwaki}, Tomoya},
        title = "{Circular polarization of gravitational waves from non-rotating supernova cores: a new probe into the pre-explosion hydrodynamics}",
      journal = {\mnras},
         year = 2018,
        month = jun,
       volume = {477},
       number = {1},
        pages = {L96-L100},
          doi = {10.1093/mnrasl/sly055},
archivePrefix = {arXiv},
       eprint = {1802.03842},
 primaryClass = {astro-ph.HE},
       adsurl = {https://ui.adsabs.harvard.edu/abs/2018MNRAS.477L..96H}
}

@ARTICLE{2018ApJ...867..126K,
       author = {{Kawahara}, Hajime and {Kuroda}, Takami and {Takiwaki}, Tomoya and {Hayama}, Kazuhiro and {Kotake}, Kei},
        title = "{A Linear and Quadratic Time-Frequency Analysis of Gravitational Waves from Core-collapse Supernovae}",
      journal = {\apj},
         year = 2018,
        month = nov,
       volume = {867},
       number = {2},
          eid = {126},
        pages = {126},
          doi = {10.3847/1538-4357/aae57b},
archivePrefix = {arXiv},
       eprint = {1810.00334},
 primaryClass = {astro-ph.HE},
       adsurl = {https://ui.adsabs.harvard.edu/abs/2018ApJ...867..126K}
}

@ARTICLE{2018ApJ...857...13P,
       author = {{Pan}, Kuo-Chuan and {Liebend{\"o}rfer}, Matthias and {Couch}, Sean M. and {Thielemann}, Friedrich-Karl},
        title = "{Equation of State Dependent Dynamics and Multi-messenger Signals from Stellar-mass Black Hole Formation}",
      journal = {\apj},
         year = 2018,
        month = apr,
       volume = {857},
       number = {1},
          eid = {13},
        pages = {13},
          doi = {10.3847/1538-4357/aab71d},
archivePrefix = {arXiv},
       eprint = {1710.01690},
 primaryClass = {astro-ph.HE},
       adsurl = {https://ui.adsabs.harvard.edu/abs/2018ApJ...857...13P}
}

@ARTICLE{2019MNRAS.482..351V,
       author = {{Vartanyan}, David and {Burrows}, Adam and {Radice}, David and {Skinner}, M. Aaron and {Dolence}, Joshua},
        title = "{A successful 3D core-collapse supernova explosion model}",
      journal = {\mnras},
         year = 2019,
        month = jan,
       volume = {482},
       number = {1},
        pages = {351-369},
          doi = {10.1093/mnras/sty2585},
archivePrefix = {arXiv},
       eprint = {1809.05106},
 primaryClass = {astro-ph.HE},
       adsurl = {https://ui.adsabs.harvard.edu/abs/2019MNRAS.482..351V}
}

@ARTICLE{2019PhRvD.100d3026S,
       author = {{Srivastava}, Varun and {Ballmer}, Stefan and {Brown}, Duncan A. and {Afle}, Chaitanya and {Burrows}, Adam and {Radice}, David and {Vartanyan}, David},
        title = "{Detection prospects of core-collapse supernovae with supernova-optimized third-generation gravitational-wave detectors}",
      journal = {\prd},
         year = 2019,
        month = aug,
       volume = {100},
       number = {4},
          eid = {043026},
        pages = {043026},
          doi = {10.1103/PhysRevD.100.043026},
archivePrefix = {arXiv},
       eprint = {1906.00084},
 primaryClass = {gr-qc},
       adsurl = {https://ui.adsabs.harvard.edu/abs/2019PhRvD.100d3026S}
}

@ARTICLE{2020MNRAS.493L.138S,
       author = {{Shibagaki}, Shota and {Kuroda}, Takami and {Kotake}, Kei and {Takiwaki}, Tomoya},
        title = "{A new gravitational-wave signature of low-T/|W| instability in rapidly rotating stellar core collapse}",
      journal = {\mnras},
         year = 2020,
        month = mar,
       volume = {493},
       number = {1},
        pages = {L138-L142},
          doi = {10.1093/mnrasl/slaa021},
archivePrefix = {arXiv},
       eprint = {1909.09730},
 primaryClass = {astro-ph.HE},
       adsurl = {https://ui.adsabs.harvard.edu/abs/2020MNRAS.493L.138S}
}

@ARTICLE{2020ApJ...898..139W,
       author = {{Warren}, MacKenzie L. and {Couch}, Sean M. and {O'Connor}, Evan P. and {Morozova}, Viktoriya},
        title = "{Constraining Properties of the Next Nearby Core-collapse Supernova with Multimessenger Signals}",
      journal = {\apj},
         year = 2020,
        month = aug,
       volume = {898},
       number = {2},
          eid = {139},
        pages = {139},
          doi = {10.3847/1538-4357/ab97b7},
archivePrefix = {arXiv},
       eprint = {1912.03328},
 primaryClass = {astro-ph.HE},
       adsurl = {https://ui.adsabs.harvard.edu/abs/2020ApJ...898..139W}
}

@ARTICLE{2020ApJ...901..108V,
       author = {{Vartanyan}, David and {Burrows}, Adam},
        title = "{Gravitational Waves from Neutrino Emission Asymmetries in Core-collapse Supernovae}",
      journal = {\apj},
         year = 2020,
        month = oct,
       volume = {901},
       number = {2},
          eid = {108},
        pages = {108},
          doi = {10.3847/1538-4357/abafac},
archivePrefix = {arXiv},
       eprint = {2007.07261},
 primaryClass = {astro-ph.HE},
       adsurl = {https://ui.adsabs.harvard.edu/abs/2020ApJ...901..108V}
}

@ARTICLE{2021ApJ...923..201E,
       author = {{Eggenberger Andersen}, Oliver and {Zha}, Shuai and {da Silva Schneider}, Andr{\'e} and {Betranhandy}, Aurore and {Couch}, Sean M. and {O'Connor}, Evan P.},
        title = "{Equation-of-state Dependence of Gravitational Waves in Core-collapse Supernovae}",
      journal = {\apj},
         year = 2021,
        month = dec,
       volume = {923},
       number = {2},
          eid = {201},
        pages = {201},
          doi = {10.3847/1538-4357/ac294c},
archivePrefix = {arXiv},
       eprint = {2106.09734},
 primaryClass = {astro-ph.HE},
       adsurl = {https://ui.adsabs.harvard.edu/abs/2021ApJ...923..201E}
}

@ARTICLE{2021ApJ...914...80P,
       author = {{Pajkos}, Michael A. and {Warren}, MacKenzie L. and {Couch}, Sean M. and {O'Connor}, Evan P. and {Pan}, Kuo-Chuan},
        title = "{Determining the Structure of Rotating Massive Stellar Cores with Gravitational Waves}",
      journal = {\apj},
         year = 2021,
        month = jun,
       volume = {914},
       number = {2},
          eid = {80},
        pages = {80},
          doi = {10.3847/1538-4357/abfb65},
archivePrefix = {arXiv},
       eprint = {2011.09000},
 primaryClass = {astro-ph.HE},
       adsurl = {https://ui.adsabs.harvard.edu/abs/2021ApJ...914...80P}
}

@ARTICLE{2021MNRAS.502.3066S,
       author = {{Shibagaki}, Shota and {Kuroda}, Takami and {Kotake}, Kei and {Takiwaki}, Tomoya},
        title = "{Characteristic time variability of gravitational-wave and neutrino signals from three-dimensional simulations of non-rotating and rapidly rotating stellar core collapse}",
      journal = {\mnras},
         year = 2021,
        month = apr,
       volume = {502},
       number = {2},
        pages = {3066-3084},
          doi = {10.1093/mnras/stab228},
archivePrefix = {arXiv},
       eprint = {2010.03882},
 primaryClass = {astro-ph.HE},
       adsurl = {https://ui.adsabs.harvard.edu/abs/2021MNRAS.502.3066S}
}

@ARTICLE{2021ApJ...914..140P,
       author = {{Pan}, Kuo-Chuan and {Liebend{\"o}rfer}, Matthias and {Couch}, Sean M. and {Thielemann}, Friedrich-Karl},
        title = "{Stellar Mass Black Hole Formation and Multimessenger Signals from Three-dimensional Rotating Core-collapse Supernova Simulations}",
      journal = {\apj},
         year = 2021,
        month = jun,
       volume = {914},
       number = {2},
          eid = {140},
        pages = {140},
          doi = {10.3847/1538-4357/abfb05},
archivePrefix = {arXiv},
       eprint = {2010.02453},
 primaryClass = {astro-ph.HE},
       adsurl = {https://ui.adsabs.harvard.edu/abs/2021ApJ...914..140P}
}

@ARTICLE{2022ApJ...924...38K,
       author = {{Kuroda}, Takami and {Fischer}, Tobias and {Takiwaki}, Tomoya and {Kotake}, Kei},
        title = "{Core-collapse Supernova Simulations and the Formation of Neutron Stars, Hybrid Stars, and Black Holes}",
      journal = {\apj},
         year = 2022,
        month = jan,
       volume = {924},
       number = {1},
          eid = {38},
        pages = {38},
          doi = {10.3847/1538-4357/ac31a8},
archivePrefix = {arXiv},
       eprint = {2109.01508},
 primaryClass = {astro-ph.HE},
       adsurl = {https://ui.adsabs.harvard.edu/abs/2022ApJ...924...38K}
}

@ARTICLE{2022MNRAS.516.1752M,
       author = {{Matsumoto}, J. and {Asahina}, Y. and {Takiwaki}, T. and {Kotake}, K. and {Takahashi}, H.~R.},
        title = "{Magnetic support for neutrino-driven explosion of 3D non-rotating core-collapse supernova models}",
      journal = {\mnras},
         year = 2022,
        month = oct,
       volume = {516},
       number = {2},
        pages = {1752-1767},
          doi = {10.1093/mnras/stac2335},
archivePrefix = {arXiv},
       eprint = {2202.07967},
 primaryClass = {astro-ph.HE},
       adsurl = {https://ui.adsabs.harvard.edu/abs/2022MNRAS.516.1752M}
}

@ARTICLE{2022MNRAS.514.3941N,
       author = {{Nakamura}, Ko and {Takiwaki}, Tomoya and {Kotake}, Kei},
        title = "{Three-dimensional simulation of a core-collapse supernova for a binary star progenitor of SN 1987A}",
      journal = {\mnras},
         year = 2022,
        month = aug,
       volume = {514},
       number = {3},
        pages = {3941-3952},
          doi = {10.1093/mnras/stac1586},
archivePrefix = {arXiv},
       eprint = {2202.06295},
 primaryClass = {astro-ph.HE},
       adsurl = {https://ui.adsabs.harvard.edu/abs/2022MNRAS.514.3941N}
}

@ARTICLE{2022PhRvD.105f3018P,
       author = {{Powell}, Jade and {M{\"u}ller}, Bernhard},
        title = "{Inferring astrophysical parameters of core-collapse supernovae from their gravitational-wave emission}",
      journal = {\prd},
         year = 2022,
        month = mar,
       volume = {105},
       number = {6},
          eid = {063018},
        pages = {063018},
          doi = {10.1103/PhysRevD.105.063018},
archivePrefix = {arXiv},
       eprint = {2201.01397},
 primaryClass = {astro-ph.HE},
       adsurl = {https://ui.adsabs.harvard.edu/abs/2022PhRvD.105f3018P}
}

@ARTICLE{2023MNRAS.520.5622B,
       author = {{Bugli}, M. and {Guilet}, J. and {Foglizzo}, T. and {Obergaulinger}, M.},
        title = "{Three-dimensional core-collapse supernovae with complex magnetic structures - II. Rotational instabilities and multimessenger signatures}",
      journal = {\mnras},
         year = 2023,
        month = apr,
       volume = {520},
       number = {4},
        pages = {5622-5634},
          doi = {10.1093/mnras/stad496},
archivePrefix = {arXiv},
       eprint = {2210.05012},
 primaryClass = {astro-ph.HE},
       adsurl = {https://ui.adsabs.harvard.edu/abs/2023MNRAS.520.5622B}
}

@ARTICLE{2023PhRvD.108j3036D,
       author = {{Drago}, Marco and {Andresen}, Haakon and {Di Palma}, Irene and {Tamborra}, Irene and {Torres-Forn{\'e}}, Alejandro},
        title = "{Multimessenger observations of core-collapse supernovae: Exploiting the standing accretion shock instability}",
      journal = {\prd},
         year = 2023,
        month = nov,
       volume = {108},
       number = {10},
          eid = {103036},
        pages = {103036},
          doi = {10.1103/PhysRevD.108.103036},
archivePrefix = {arXiv},
       eprint = {2305.07688},
 primaryClass = {astro-ph.HE},
       adsurl = {https://ui.adsabs.harvard.edu/abs/2023PhRvD.108j3036D}
}

@ARTICLE{2023PhRvL.131s1201J,
       author = {{Jakobus}, Pia and {M{\"u}ller}, Bernhard and {Heger}, Alexander and {Zha}, Shuai and {Powell}, Jade and {Motornenko}, Anton and {Steinheimer}, Jan and {St{\"o}cker}, Horst},
        title = "{Gravitational Waves from a Core g Mode in Supernovae as Probes of the High-Density Equation of State}",
      journal = {\prl},
         year = 2023,
        month = nov,
       volume = {131},
       number = {19},
          eid = {191201},
        pages = {191201},
          doi = {10.1103/PhysRevLett.131.191201},
archivePrefix = {arXiv},
       eprint = {2301.06515},
 primaryClass = {astro-ph.HE},
       adsurl = {https://ui.adsabs.harvard.edu/abs/2023PhRvL.131s1201J}
}

@ARTICLE{2023PhRvD.107j3025K,
       author = {{Kuroda}, Takami and {Shibata}, Masaru},
        title = "{Spontaneous scalarization as a new core-collapse supernova mechanism and its multimessenger signals}",
      journal = {\prd},
         year = 2023,
        month = may,
       volume = {107},
       number = {10},
          eid = {103025},
        pages = {103025},
          doi = {10.1103/PhysRevD.107.103025},
archivePrefix = {arXiv},
       eprint = {2302.09853},
 primaryClass = {astro-ph.HE},
       adsurl = {https://ui.adsabs.harvard.edu/abs/2023PhRvD.107j3025K}
}

@ARTICLE{2023ApJ...959...21P,
       author = {{Pajkos}, Michael A. and {VanCamp}, Steven J. and {Pan}, Kuo-Chuan and {Vartanyan}, David and {Deppe}, Nils and {Couch}, Sean M.},
        title = "{Characterizing the Directionality of Gravitational Wave Emission from Matter Motions within Core-collapse Supernovae}",
      journal = {\apj},
         year = 2023,
        month = dec,
       volume = {959},
       number = {1},
          eid = {21},
        pages = {21},
          doi = {10.3847/1538-4357/acfca4},
archivePrefix = {arXiv},
       eprint = {2306.01919},
 primaryClass = {astro-ph.HE},
       adsurl = {https://ui.adsabs.harvard.edu/abs/2023ApJ...959...21P}
}

@ARTICLE{2023PhRvD.107j3015V,
       author = {{Vartanyan}, David and {Burrows}, Adam and {Wang}, Tianshu and {Coleman}, Matthew S.~B. and {White}, Christopher J.},
        title = "{Gravitational-wave signature of core-collapse supernovae}",
      journal = {\prd},
         year = 2023,
        month = may,
       volume = {107},
       number = {10},
          eid = {103015},
        pages = {103015},
          doi = {10.1103/PhysRevD.107.103015},
archivePrefix = {arXiv},
       eprint = {2302.07092},
 primaryClass = {astro-ph.HE},
       adsurl = {https://ui.adsabs.harvard.edu/abs/2023PhRvD.107j3015V}
}

@ARTICLE{2023PhRvD.107l3005A,
       author = {{Afle}, Chaitanya and {Kundu}, Suman Kumar and {Cammerino}, Jenna and {Coughlin}, Eric R. and {Brown}, Duncan A. and {Vartanyan}, David and {Burrows}, Adam},
        title = "{Measuring the properties of f -mode oscillations of a protoneutron star by third-generation gravitational-wave detectors}",
      journal = {\prd},
         year = 2023,
        month = jun,
       volume = {107},
       number = {12},
          eid = {123005},
        pages = {123005},
          doi = {10.1103/PhysRevD.107.123005},
archivePrefix = {arXiv},
       eprint = {2304.04283},
 primaryClass = {astro-ph.IM},
       adsurl = {https://ui.adsabs.harvard.edu/abs/2023PhRvD.107l3005A}
}

@ARTICLE{2023PhRvD.107h3029B,
       author = {{Bruel}, Tristan and {Bizouard}, Marie-Anne and {Obergaulinger}, Martin and {Maturana-Russel}, Patricio and {Torres-Forn{\'e}}, Alejandro and {Cerd{\'a}-Dur{\'a}n}, Pablo and {Christensen}, Nelson and {Font}, Jos{\'e} A. and {Meyer}, Renate},
        title = "{Inference of protoneutron star properties in core-collapse supernovae from a gravitational-wave detector network}",
      journal = {\prd},
         year = 2023,
        month = apr,
       volume = {107},
       number = {8},
          eid = {083029},
        pages = {083029},
          doi = {10.1103/PhysRevD.107.083029},
archivePrefix = {arXiv},
       eprint = {2301.10019},
 primaryClass = {astro-ph.HE},
       adsurl = {https://ui.adsabs.harvard.edu/abs/2023PhRvD.107h3029B}
}

@ARTICLE{2023PhRvD.107h3017L,
       author = {{Lin}, Zidu and {Rijal}, Abhinav and {Lunardini}, Cecilia and {Morales}, Manuel D. and {Zanolin}, Michele},
        title = "{Characterizing a supernova's standing accretion shock instability with neutrinos and gravitational waves}",
      journal = {\prd},
         year = 2023,
        month = apr,
       volume = {107},
       number = {8},
          eid = {083017},
        pages = {083017},
          doi = {10.1103/PhysRevD.107.083017},
archivePrefix = {arXiv},
       eprint = {2211.07878},
 primaryClass = {astro-ph.HE},
       adsurl = {https://ui.adsabs.harvard.edu/abs/2023PhRvD.107h3017L}
}

@ARTICLE{2024arXiv240513211G,
       author = {{Gill}, Kiranjyot},
        title = "{Milli-to-Deci-Hertz Detection Prospects for Gravitational Waves from Core-Collapse Supernovae}",
      journal = {arXiv e-prints},
         year = 2024,
        month = may,
          eid = {arXiv:2405.13211},
        pages = {arXiv:2405.13211},
          doi = {10.48550/arXiv.2405.13211},
archivePrefix = {arXiv},
       eprint = {2405.13211},
 primaryClass = {astro-ph.HE},
       adsurl = {https://ui.adsabs.harvard.edu/abs/2024arXiv240513211G}
}

@ARTICLE{2024ApJ...975...12C,
       author = {{Choi}, Lyla and {Burrows}, Adam and {Vartanyan}, David},
        title = "{Gravitational-wave and Gravitational-wave Memory Signatures of Core-collapse Supernovae}",
      journal = {\apj},
         year = 2024,
        month = nov,
       volume = {975},
       number = {1},
          eid = {12},
        pages = {12},
          doi = {10.3847/1538-4357/ad74f8},
archivePrefix = {arXiv},
       eprint = {2408.01525},
 primaryClass = {astro-ph.HE},
       adsurl = {https://ui.adsabs.harvard.edu/abs/2024ApJ...975...12C}
}

@ARTICLE{1978Natur.274..565T,
       author = {{Turner}, M.~S.},
        title = "{Gravitational radiation from supernova neutrino bursts}",
      journal = {\nat},
         year = 1978,
        month = aug,
       volume = {274},
       number = {5671},
        pages = {565-566},
          doi = {10.1038/274565a0},
       adsurl = {https://ui.adsabs.harvard.edu/abs/1978Natur.274..565T}
}

@ARTICLE{2010CQGra..27s4005Y,
       author = {{Yakunin}, Konstantin N. and {Marronetti}, Pedro and {Mezzacappa}, Anthony and {Bruenn}, Stephen W. and {Lee}, Ching-Tsai and {Chertkow}, Merek A. and {Hix}, W. Raphael and {Blondin}, John M. and {Lentz}, Eric J. and {Messer}, O.~E. Bronson and {Yoshida}, Shin'ichirou},
        title = "{Gravitational waves from core collapse supernovae}",
      journal = {Classical and Quantum Gravity},
         year = 2010,
        month = oct,
       volume = {27},
       number = {19},
          eid = {194005},
        pages = {194005},
          doi = {10.1088/0264-9381/27/19/194005},
archivePrefix = {arXiv},
       eprint = {1005.0779},
 primaryClass = {gr-qc},
       adsurl = {https://ui.adsabs.harvard.edu/abs/2010CQGra..27s4005Y}
}

@ARTICLE{2011ApJ...743...30T,
       author = {{Takiwaki}, Tomoya and {Kotake}, Kei},
        title = "{Gravitational Wave Signatures of Magnetohydrodynamically Driven Core-collapse Supernova Explosions}",
      journal = {\apj},
         year = 2011,
        month = dec,
       volume = {743},
       number = {1},
          eid = {30},
        pages = {30},
          doi = {10.1088/0004-637X/743/1/30},
archivePrefix = {arXiv},
       eprint = {1004.2896},
 primaryClass = {astro-ph.HE},
       adsurl = {https://ui.adsabs.harvard.edu/abs/2011ApJ...743...30T}
}

@ARTICLE{2022MNRAS.510.5535J,
       author = {{Jardine}, Rylan and {Powell}, Jade and {M{\"u}ller}, Bernhard},
        title = "{Gravitational wave signals from 2D core-collapse supernova models with rotation and magnetic fields}",
      journal = {\mnras},
         year = 2022,
        month = mar,
       volume = {510},
       number = {4},
        pages = {5535-5552},
          doi = {10.1093/mnras/stab3763},
archivePrefix = {arXiv},
       eprint = {2105.01315},
 primaryClass = {astro-ph.HE},
       adsurl = {https://ui.adsabs.harvard.edu/abs/2022MNRAS.510.5535J}
}

@ARTICLE{2023MNRAS.522.6070P,
       author = {{Powell}, Jade and {M{\"u}ller}, Bernhard and {Aguilera-Dena}, David R. and {Langer}, Norbert},
        title = "{Three dimensional magnetorotational core-collapse supernova explosions of a 39 solar mass progenitor star}",
      journal = {\mnras},
         year = 2023,
        month = jul,
       volume = {522},
       number = {4},
        pages = {6070-6086},
          doi = {10.1093/mnras/stad1292},
archivePrefix = {arXiv},
       eprint = {2212.00200},
 primaryClass = {astro-ph.HE},
       adsurl = {https://ui.adsabs.harvard.edu/abs/2023MNRAS.522.6070P}
}

@ARTICLE{2021arXiv210505862M,
       author = {{Mukhopadhyay}, Mainak and {Cardona}, Carlos and {Lunardini}, Cecilia},
        title = "{The neutrino gravitational memory from a core collapse supernova: phenomenology and physics potential}",
      journal = {arXiv e-prints},
         year = 2021,
        month = may,
          eid = {arXiv:2105.05862},
        pages = {arXiv:2105.05862},
          doi = {10.48550/arXiv.2105.05862},
archivePrefix = {arXiv},
       eprint = {2105.05862},
 primaryClass = {astro-ph.HE},
       adsurl = {https://ui.adsabs.harvard.edu/abs/2021arXiv210505862M}
}

@ARTICLE{2022PhRvD.106d3020M,
       author = {{Mukhopadhyay}, Mainak and {Lin}, Zidu and {Lunardini}, Cecilia},
        title = "{Memory-triggered supernova neutrino detection}",
      journal = {\prd},
         year = 2022,
        month = aug,
       volume = {106},
       number = {4},
          eid = {043020},
        pages = {043020},
          doi = {10.1103/PhysRevD.106.043020},
archivePrefix = {arXiv},
       eprint = {2110.14657},
 primaryClass = {astro-ph.HE},
       adsurl = {https://ui.adsabs.harvard.edu/abs/2022PhRvD.106d3020M}
}

@INPROCEEDINGS{2019BAAS...51g..77T,
       author = {{Thorpe}, James Ira and {Ziemer}, John and {Thorpe}, Ira and {Livas}, Jeff and {Conklin}, John W. and {Caldwell}, Robert and {Berti}, Emanuele and {McWilliams}, Sean T. and {Stebbins}, Robin and {Shoemaker}, David and {Ferrara}, Elizabeth C. and {Larson}, Shane L. and {Shoemaker}, Deirdre and {Key}, Joey Shapiro and {Vallisneri}, Michele and {Eracleous}, Michael and {Schnittman}, Jeremy and {Kamai}, Brittany and {Camp}, Jordan and {Mueller}, Guido and {Bellovary}, Jillian and {Rioux}, Norman and {Baker}, John and {Bender}, Peter L. and {Cutler}, Curt and {Cornish}, Neil and {Hogan}, Craig and {Manthripragada}, Sridhar and {Ware}, Brent and {Natarajan}, Priyamvada and {Numata}, Kenji and {Sankar}, Shannon R. and {Kelly}, Bernard J. and {McKenzie}, Kirk and {Slutsky}, Jacob and {Spero}, Robert and {Hewitson}, Martin and {Francis}, Samuel and {DeRosa}, Ryan and {Yu}, Anthony and {Hornschemeier}, Ann and {Wass}, Peter},
        title = "{The Laser Interferometer Space Antenna: Unveiling the Millihertz Gravitational Wave Sky}",
    booktitle = {Bulletin of the American Astronomical Society},
         year = 2019,
       volume = {51},
        month = sep,
          eid = {77},
        pages = {77},
          doi = {10.48550/arXiv.1907.06482},
archivePrefix = {arXiv},
       eprint = {1907.06482},
 primaryClass = {astro-ph.IM},
       adsurl = {https://ui.adsabs.harvard.edu/abs/2019BAAS...51g..77T}
}

@INPROCEEDINGS{2019BAAS...51g..35R,
       author = {{Reitze}, David and {Adhikari}, Rana X. and {Ballmer}, Stefan and {Barish}, Barry and {Barsotti}, Lisa and {Billingsley}, GariLynn and {Brown}, Duncan A. and {Chen}, Yanbei and {Coyne}, Dennis and {Eisenstein}, Robert and {Evans}, Matthew and {Fritschel}, Peter and {Hall}, Evan D. and {Lazzarini}, Albert and {Lovelace}, Geoffrey and {Read}, Jocelyn and {Sathyaprakash}, B.~S. and {Shoemaker}, David and {Smith}, Joshua and {Torrie}, Calum and {Vitale}, Salvatore and {Weiss}, Rainer and {Wipf}, Christopher and {Zucker}, Michael},
        title = "{Cosmic Explorer: The U.S. Contribution to Gravitational-Wave Astronomy beyond LIGO}",
    booktitle = {Bulletin of the American Astronomical Society},
         year = 2019,
       volume = {51},
        month = sep,
          eid = {35},
        pages = {35},
          doi = {10.48550/arXiv.1907.04833},
archivePrefix = {arXiv},
       eprint = {1907.04833},
 primaryClass = {astro-ph.IM},
       adsurl = {https://ui.adsabs.harvard.edu/abs/2019BAAS...51g..35R}
}

@ARTICLE{2020JCAP...03..050M,
       author = {{Maggiore}, Michele and {Van Den Broeck}, Chris and {Bartolo}, Nicola and {Belgacem}, Enis and {Bertacca}, Daniele and {Bizouard}, Marie Anne and {Branchesi}, Marica and {Clesse}, Sebastien and {Foffa}, Stefano and {Garc{\'\i}a-Bellido}, Juan and {Grimm}, Stefan and {Harms}, Jan and {Hinderer}, Tanja and {Matarrese}, Sabino and {Palomba}, Cristiano and {Peloso}, Marco and {Ricciardone}, Angelo and {Sakellariadou}, Mairi},
        title = "{Science case for the Einstein telescope}",
      journal = {\jcap},
         year = 2020,
        month = mar,
       volume = {2020},
       number = {3},
          eid = {050},
        pages = {050},
          doi = {10.1088/1475-7516/2020/03/050},
archivePrefix = {arXiv},
       eprint = {1912.02622},
 primaryClass = {astro-ph.CO},
       adsurl = {https://ui.adsabs.harvard.edu/abs/2020JCAP...03..050M}
}

@ARTICLE{2016PhRvD..93d2004K,
       author = {{Klimenko}, S. and {Vedovato}, G. and {Drago}, M. and {Salemi}, F. and {Tiwari}, V. and {Prodi}, G.~A. and {Lazzaro}, C. and {Ackley}, K. and {Tiwari}, S. and {Da Silva}, C.~F. and {Mitselmakher}, G.},
        title = "{Method for detection and reconstruction of gravitational wave transients with networks of advanced detectors}",
      journal = {\prd},
         year = 2016,
        month = feb,
       volume = {93},
       number = {4},
          eid = {042004},
        pages = {042004},
          doi = {10.1103/PhysRevD.93.042004},
archivePrefix = {arXiv},
       eprint = {1511.05999},
 primaryClass = {gr-qc},
       adsurl = {https://ui.adsabs.harvard.edu/abs/2016PhRvD..93d2004K}
}

@software{2021zndo...5798976K,
       author = {{Klimenko}, Sergey and {Vedovato}, Gabriele and {Necula}, Valentin and {Salemi}, Francesco and {Drago}, Marco and {Poulton}, Rhys and {Chassande-Mottin}, Eric and {Tiwari}, Vaibhav and {Lazzaro}, Claudia and {O'Brian}, Brendan and {Szczepanczyk}, Marek and {Tiwari}, Shubhanshu and {Gayathri}, V.},
        title = "{cWB pipeline library: 6.4.1}",
         year = 2021,
        month = dec,
          eid = {10.5281/zenodo.5798976},
          doi = {10.5281/zenodo.5798976},
      version = {cWB-6.4.1},
    publisher = {Zenodo},
       adsurl = {https://ui.adsabs.harvard.edu/abs/2021zndo...5798976K}
}

@ARTICLE{2024PhRvD.110d2007S,
       author = {{Szczepa{\'n}czyk}, Marek J. and {Zheng}, Yanyan and {Antelis}, Javier M. and {Benjamin}, Michael and {Bizouard}, Marie-Anne and {Casallas-Lagos}, Alejandro and {Cerd{\'a}-Dur{\'a}n}, Pablo and {Davis}, Derek and {Gondek-Rosi{\'n}ska}, Dorota and {Klimenko}, Sergey and {Moreno}, Claudia and {Obergaulinger}, Martin and {Powell}, Jade and {Ramirez}, Dymetris and {Ratto}, Brad and {Richardson}, Colter and {Rijal}, Abhinav and {Stuver}, Amber L. and {Szewczyk}, Pawe{\l} and {Vedovato}, Gabriele and {Zanolin}, Michele and {Bartos}, Imre and {Bhaumik}, Shubhagata and {Bulik}, Tomasz and {Drago}, Marco and {Font}, Jos{\'e} A. and {De Colle}, Fabio and {Garc{\'\i}a-Bellido}, Juan and {Gayathri}, V. and {Hughey}, Brennan and {Mitselmakher}, Guenakh and {Mishra}, Tanmaya and {Mukherjee}, Soma and {Nguyen}, Quynh Lan and {Chan}, Man Leong and {Di Palma}, Irene and {Piotrzkowski}, Brandon J. and {Singh}, Neha},
        title = "{Optically targeted search for gravitational waves emitted by core-collapse supernovae during the third observing run of Advanced LIGO and Advanced Virgo}",
      journal = {\prd},
         year = 2024,
        month = aug,
       volume = {110},
       number = {4},
          eid = {042007},
        pages = {042007},
          doi = {10.1103/PhysRevD.110.042007},
archivePrefix = {arXiv},
       eprint = {2305.16146},
 primaryClass = {astro-ph.HE},
       adsurl = {https://ui.adsabs.harvard.edu/abs/2024PhRvD.110d2007S}
}

@ARTICLE{2006RPPh...69..971K,
       author = {{Kotake}, Kei and {Sato}, Katsuhiko and {Takahashi}, Keitaro},
        title = "{Explosion mechanism, neutrino burst and gravitational wave in core-collapse supernovae}",
      journal = {Reports on Progress in Physics},
         year = 2006,
        month = apr,
       volume = {69},
       number = {4},
        pages = {971-1143},
          doi = {10.1088/0034-4885/69/4/R03},
archivePrefix = {arXiv},
       eprint = {astro-ph/0509456},
 primaryClass = {astro-ph},
       adsurl = {https://ui.adsabs.harvard.edu/abs/2006RPPh...69..971K}
}

@ARTICLE{2025Sci...389.1012L,
       author = {{Buchli}, Jonas and {Tracey}, Brendan and {Andric}, Tomislav and {Wipf}, Christopher and {Chiu}, Yu Him Justin and {Lochbrunner}, Matthias and {Donner}, Craig and {Adhikari}, Rana X. and {Harms}, Jan and {Barr}, Iain and {Hafner}, Roland and {Huber}, Andrea and {Abdolmaleki}, Abbas and {Beattie}, Charlie and {Betzwieser}, Joseph and {Cabi}, Serkan and {Degrave}, Jonas and {Dong}, Yuzhu and {Fritz}, Leslie and {Gupta}, Anchal and {Groth}, Oliver and {Huang}, Sandy and {Norman}, Tamara and {Openshaw}, Hannah and {Rollins}, Jameson and {Thornton}, Greg and {van den Driessche}, George and {Wulfmeier}, Markus and {Kohli}, Pushmeet and {Riedmiller}, Martin and {LIGO Instrument Team}},
        title = "{Improving cosmological reach of a gravitational wave observatory using Deep Loop Shaping}",
      journal = {Science},
         year = 2025,
        month = sep,
       volume = {389},
       number = {6764},
        pages = {1012-1015},
          doi = {10.1126/science.adw1291},
archivePrefix = {arXiv},
       eprint = {2509.14016},
 primaryClass = {astro-ph.IM},
       adsurl = {https://ui.adsabs.harvard.edu/abs/2025Sci...389.1012L}
}

@ARTICLE{1996PhRvL..76..352B,
       author = {{Burrows}, Adam and {Hayes}, John},
        title = "{Pulsar Recoil and Gravitational Radiation Due to Asymmetrical Stellar Collapse and Explosion}",
      journal = {\prl},
         year = 1996,
        month = jan,
       volume = {76},
       number = {3},
        pages = {352-355},
          doi = {10.1103/PhysRevLett.76.352},
archivePrefix = {arXiv},
       eprint = {astro-ph/9511106},
 primaryClass = {astro-ph},
       adsurl = {https://ui.adsabs.harvard.edu/abs/1996PhRvL..76..352B}
}

@ARTICLE{2026PhRvD.113h3034L,
       author = {{Lella}, Alessandro and {Lucente}, Giuseppe and {Kresse}, Daniel and {Glas}, Robert and {Janka}, Hans-Thomas and {Mirizzi}, Alessandro},
        title = "{Gravitational-wave signals for supernova explosions of three-dimensional progenitors}",
      journal = {\prd},
         year = 2026,
        month = apr,
       volume = {113},
       number = {8},
          eid = {083034},
        pages = {083034},
          doi = {10.1103/f3n4-k2cq},
archivePrefix = {arXiv},
       eprint = {2602.02651},
 primaryClass = {astro-ph.HE},
       adsurl = {https://ui.adsabs.harvard.edu/abs/2026PhRvD.113h3034L}
}

@STRING{aap = "Astron. Astrophys."}

@STRING{apjl = "Ap.J."}

@STRING{apjs = "Ap.J. Suppl."}

@STRING{apj = "Ap.J."}

@STRING{mnras = "Mon. Not. R. Ast. Soc."}

@STRING{nat = "Nature"}

@STRING{science = "Science"}

@STRING{st = "S\&T"}

@STRING{as = "Appl. Spectrosc."}

@STRING{geo = "Geophysics"}

@STRING{on = "Opt. News"}

@STRING{prd = "Phys. Rev. D"}

@STRING{prl = "Phys. Rev. Lett."}

@STRING{st = "Sidereal Times"}

@STRING{sky = "Sky Telesc."}

@STRING{space = "Space"}

@STRING{aap = aa}

@STRING{apjs = apjsup}

@STRING{apjl = apj}

@STRING{mnras ="Mon. Not. Roy. Ast. Soc."}

\end{document}